\documentclass[PCJ,Unicode,screen]{cedram}

\usepackage[natbib=true,mincitenames=1,maxcitenames=2,maxbibnames=20,minbibnames=20,uniquelist=false,uniquename=false,style=pcj]{biblatex}
\usepackage{lipsum}
\usepackage{setspace}
\usepackage{booktabs}
\usepackage{multirow}
\usepackage{tikz}

\providecommand*{\noopsort}[1]{}

\DeclareUnicodeCharacter{202F}{\,}
\DOI{10.48550/arXiv.2408.16269} % this is the DOI of the preprint on the preprint server
\PCIcorresp{baptiste.sorin@outlook.com}
\title[Sample for the template]{Individual or collective treatments: how to target antimicrobial use to limit the spread of \textit{Mannheimia haemolytica} among beef cattle?}
 \author{\firstname{Baptiste} \lastname{Sorin-Dupont}}
 \address{Oniris, INRAE, BIOEPAR, 44300, Nantes, France}
 \email[B. SORIN-DUPONT]{baptiste.sorin@outlook.com}
 \author{\firstname{Antoine} \lastname{Poyard}}
   \address{Oniris, INRAE, BIOEPAR, 44300, Nantes, France}

 \author{\firstname{Sebastien} \lastname{Assié}}
  \address{Oniris, INRAE, BIOEPAR, 44300, Nantes, France}

 \author{\firstname{Sebastien} \lastname{Picault}}
  \address{Oniris, INRAE, BIOEPAR, 44300, Nantes, France}

\author{\firstname{Pauline} \lastname{Ezanno}}
 \address{Oniris, INRAE, BIOEPAR, 44300, Nantes, France}

\nolinenumbers
 \begin{abstract}
The overuse of antibiotics has become a major global concern due to its role in diminishing treatment effectiveness and positively selecting antibiotic-resistant bacterial strains. This issue is particularly important in the beef cattle sector, where Bovine Respiratory Diseases (BRD) impose significant economic and welfare burdens. BRD are complex, multifactorial conditions primarily affecting young calves and feedlot cattle, caused by a combination of viral and bacterial pathogens, environmental factors, and stressors. Despite efforts to reduce antimicrobial use (AMU), the cattle production system remains heavily reliant on antibiotics to control BRD, often through the implementation of collective treatments to prevent outbreaks.\\
This study aimed at evaluating the impact of various strategies of collective treatments with antimicrobials on the spread of a  BRD pathogen, \textit{Mannheimia haemolytica}, specifically focusing on criteria for implementing collective treatments. Using a mechanistic stochastic model, we simulated the spread of \textit{M. haemolytica} in a multi-pen fattening operation under sixteen different scenarios, considering pen composition, individual risk levels, and treatment strategies. Our findings suggest that an alternative criterion for collective treatments based on the speed of the disease spread, could reduce BRD incidence and AMU more effectively than conventional methods. This research highlights the importance of responsible collective treatment strategies and the potential benefits of novel criteria for collective treatment in improving animal health. Moreover, it emphasizes the need for transparency on the exposure to risk factors along the production chain. 
 \end{abstract}
\begin{document}

\maketitle
%you may choose to use a table of content
%\tableofcontents
\onehalfspacing
\section*{Introduction}
Excessive and inappropriate use of antimicrobials can select resistant bacterial strains, making it harder to treat infections in both animals and humans (\cite{woolums_multidrug_2018,chen_scoping_2022}). However, antimicrobial treatments remain the main control measure for health practioners to treat bacterial infectious diseases, thus making antimicrobioresistance a global concern (\cite{laxminarayan_antibiotic_2013,coetzee_association_2019}).\\

Bovine respiratory diseases (BRD) pose a significant burden on the global livestock industry, causing considerable economic losses and welfare concerns for cattle (\cite{delabouglise_linking_2017}). BRD are complex multifactorial respiratory conditions primarily affecting young calves and feedlot cattle (\cite{griffin_monster_2014}). They are caused by a combination of viral and bacterial pathogens, environmental factors, and stressors (\cite{kudirkiene_occurrence_2021}). The bovine respiratory syncytial virus (BRSV) or the parainfluenza-3 (IPV-3) are frequently reported viral agents whereas common bacteria include \textit{Pasteurella multocida}, \textit{Mannheimia haemolytica} or \textit{Mycoplasma} spp. (\cite{caswell_mycoplasma_2010,timsit_transmission_2013,grissett_structured_2015}). They lead to reduced feed efficiency, growth rates, and overall productivity, resulting in substantial financial losses for producers (\cite{griffin_bovine_2010,blakebrough-hall_factors_2022}). Additionally, they cause significant suffering in affected animals, characterized by symptoms such as coughing, fever, and respiratory distress (\cite{hilton_brd_2014}). The prevention and management of BRD are crucial for maintaining animal health, welfare, and sustainable livestock production (\cite{hilton_brd_2014}).\\

While viral pathogens are typically more infectious, much of the internal damage and clinical signs associated with BRD arise from secondary bacterial infections (\cite{rice_mannheimia_2007}). Pasteurellaceae such as \textit{Mannheimia haemolytica} or \textit{Pasteurella multocida} are often associated with severe BRD cases, based upon necropsies of animals presenting severe respiratory symptomes, as well as positive serologies (\cite{booker_microbiological_2008,welsh_isolation_2004}). \textit{M.haemolytica} is particularly recurrent, with seroprevalences up to 80\% and incidence risks up to 22.7\% (\cite{assie_exposure_2009}). \textit{M. haemolytica} is commonly reported during BRD episodes in the field (\cite{klima_characterization_2012,sudaryatma_bovine_2018}). Curative treatments for BRD mostly focus on these secondary infections, as they consist in using antimicrobials (\cite{edwards_control_2010}). \textit{M.haemolitica}'s response to antimicrobials is well characterized, making it a good candidate for a pathogen specific studies investigating treatment strategies (\cite{noyes_mannheimia_2015}).\\

Despite considerable progress made in the livestock sector to reduce antimicrobial use (AMU) (\cite{bateman_evaluation_1990,watts_antimicrobial_2010}) the cattle production system is still heavily reliant on this method to control BRD (\cite{oconnor_mixed_2013,abell_mixed_2017}). Pathogen identification is not done on routine, therefore antimicrobial curative treatment is the baseline response in the event of an outbreak (\cite{dedonder_review_2015}). Individual treatment when animals show signs of disease is mostly used to care for affected animals, but collective treatments are also widely embraced to prevent BRD outbreaks and minimize economic losses (\cite{nickell_metaphylactic_2010,ollivett_brd_2020,baggott2011demonstration}). The latter involves the administration of appropriate medications, such as antibiotics or anti-inflammatory drugs, to affected animals as well as the entire group to prevent additional cases (\cite{lees_rational_2002,ives_use_2015}). This solution is motivated by poor sensitivity of current BRD field diagnostic methods, typical pathogenesis of BRD, and labor issues (\cite{ives_use_2015}). Indeed, the sensitivity for BRD diagnosis was evaluated at around 0.27, making it very unefficient to treat each animal individually upon visual appraisal (\cite{timsit2016diagnostic}). As a result, a trade-off exists between antimicrobial use and bovine respiratory disease (BRD) occurrence. This trade-off highlights the need for responsible antimicrobial use practices, such as targeted collective treatments with antimicrobials. Yet, to this day, such strategies have not been proposed. Indeed, the relevant criteria and associated threshold to implement collective treatments are still questioned (\cite{edwards_control_2010}).\\

Modeling is used to better understand complex infection dynamics and compare scenarios and rank interventions (\cite{ezanno_how_2020}). In the case of BRD, a model ranking treatment strategies at pen level has been published (\cite{picault_modelling_2022}). It simulates the circulation of an average pathogen in conditions mimicking contrasted farming contexts such as small versus large pens (considered as isolated from other pens) and low versus high risk of contracting BRD at pen level. The authors concluded on a bigger BRD circulation in large pens composed of high risk individuals and discussed which pens could get the highest benefit from collective treatments. A second model compares the impact of farm management practices on the spread of three causal pathogens of BRD in multi-pen farms featuring several individual risk levels of contracting BRD (\cite{sorin-dupont_modeling_2023}). This study highlighted that sorting animals in pens in accordance with their individual risk level of developing BRD participates in limiting the spread of the most contagious pathogens. However, no model integrating novel criteria for triggering collective treatments has been proposed so far. \\

This study aimed at evaluating the impact of collective treatment strategies on the spread of \textit{M.haemolytica} in contrasted farming scenarios. More specifically, the objective was to evaluate criteria for implementing collective treatments in order to identify the best criterion. To that end, we used a mechanistic stochastic model of the spread of \textit{Mannheimia haemolytica} in a multi-pen fattening operation. We investigated sixteen contrasted scenarios reflecting farming practices in fattening operations, based on pen composition, individual risk level distribution, and individual vs. collective treatments  based on conventional and original criteria. We compared scenario outputs regarding case occurrence, severity, mortality and AMU. \\

\section*{Material and methods }
\subsection*{Model processes and assumptions}

\begin{figure}[h!]
    \centering
    \includegraphics[width=1\textwidth]{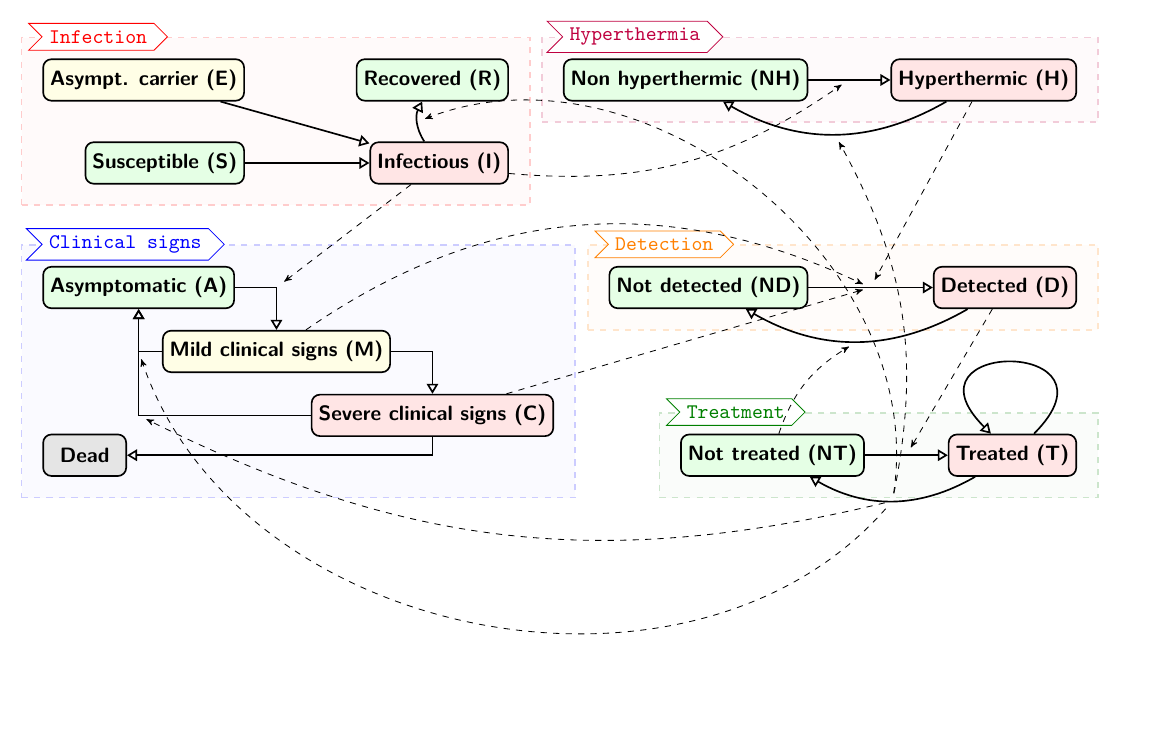}
    \caption{Graphic overview of the states and transitions in each process of the model. Each animal is defined by its state in each of the 5 state machines included in the model. Full arrows represent a transition from one state to another whereas the dashed lines represent regulatory links. Let $X$ be a regulator (state or transition) of transition  $T_2$. This will be represented by a dashed arrow starting from $X$ and targeting $T_2$.  From \cite{sorin-dupont_modeling_2023}}
    \label{graphic_overview}
\end{figure}

We modeled the spread of \textit{M. haemolytica} in a multi-pen fattening farm and the treatment of associated BRD. We extended our stochastic and mechanistic individual-based model previously published (\cite{sorin-dupont_modeling_2023}) for studying the circulation of specific pathogens involved in BRD, including \textit{M. haemolytica}, within a fattening farm building, by including the possibility of collective treatments based on conventional and original criteria.
Our model monitored 6 individual characteristics, impletemented either as input parameters or state machines. First, the model featured the individual risk status of developing BRD as an input parameter, which corresponded to an individual qualitative information (low, medium, high) summarizing the level of stressors predisposing to BRD that the animal had been exposed to prior to its arrival. The model also featured state machines defining hyperthermia (either due to BRD or unspecific), infection status (susceptible, asymptomatic carrier, infectious), clinical status (asymptomatic, or with mild or severe clinical signs), detection status (detected or not), and treatment status (treated or not treated). Hereafter, we detail how treatments are triggered in this new version of the model. A graphic overview of the model state machines is proposed in Figure \ref{graphic_overview}. \\

\subsubsection*{Hyperthermia}
Hyperthermia was composed of two states: hyperthermic (H) and non-hyperthermic (NH). NH animals could transition to H with probability $p_H$ attributed to non-infectious factors. Once in state H, they remained there for a period $\tau_H$ sampled from a Beta distribution adjusted using observed data, before reverting to NH. Additionally, the transitions from NH to H and back could result from the infection process with a 1 probability.

\subsubsection*{Infection and clinical signs status}
Four infection statuses were considered: susceptible (S), asymptomatic carrier (E), infectious (I) and recovered (R) animals. Asymptomatic carriers were modeled as non infectious animals which carry \textit{M.haemolytica} commensally. They could spontaneously turn I with probability $p_E$ and could also be infected by surrounding infectious individuals (I). Three actions were triggered when entering the state I: (1) the individual exhibited mild clinical signs for a duration $\tau_M$ drawn from a Beta distribution calibrated from observed data, (2) the animal changed from NH to H state,  (3) a random draw with probability $p_C$ drove whether the individual would display severe clinical signs at the end of its mild clinical signs. The probability $p_C$ is higher if the animal has a high individual risk level. If displaying severe clinical signs, a boolean deciding on the survival of the individual was drawn from a binomial law of probability $p_d$. Death occurred at the end of the severe clinical signs duration ($\tau_C$). If the individual did not die and was still infectious after $\tau_C$, it transitioned back to MildC. The animal then recovered and became R at the end of the infectious period. Recovery occurred after a duration $\tau_I$ drawn from a gamma distribution. Theoretically, $\tau_I$  is longer than $\tau_M$. However, $\tau_M+ \tau_C $ could exceed $\tau_I$. In that case, the infectious period was $\tau_M+ \tau_C $. When transitioning to R state, animals changed from H to NH. \\

\subsubsection*{Detection status}
Similarly to previous models, animals could be detected upon exhibition of clinical signs or hyperthermia. Severe clinical signs were detected with a sensitivity of 0.6, while the sensitivity for mild clinical signs detection was averaged to 0.3, assuming that lethargy is the main driver for detection of severe clinical signs (\cite{kayser_evaluation_2019}). We assumed a clinical check-up at every time step (12 h, in agreement with farmers' herd monitoring frequency (\cite{terry_strategies_2021}). After detecting the first case through visual appraisal, all hyperthermic animals were identified using rectal temperature measured at the next feeding time, 12 hours later. (\cite{picault_modelling_2022,sorin-dupont_modeling_2023}). False discovery may occur with the temperature check as hyperthermia may be unrelated to infection.

\subsubsection*{Treatment status} The treatment process represented a standard antimicrobial treatment protocol during the fattening period.  Each animal that was detected as affected, either though visual detection or hyperthermia, received one antibiotic dose, assumed to be effective after duration $\tau_T$ with a probability $p_T$, accounting for antimicrobial resistance. Animals still exhibiting clinical signs after this duration were treated again (same duration), the number of treatments per episode being limited ($max_T$). Transitions from T to NT occurred in three cases: (1) the animal recovered after duration $\tau_T$ due to the treatment success with probability $p_T$; (2) the end of the infectious period occurred while under treatment but was not caused by it; (3) the animal still expressed clinical signs but was already treated $max_T$  times, thus was not treated again and ignored from further treatments, as per on-field recommendations (\cite{van2017reducing}). The values of the parameters as well as their source are summarized in Table \ref{table_mh}. \\
\newpage

\begin{table}[h!]
\centering
\begin{tabular}{@{}p{5.5cm}p{1.5cm}p{7cm}p{2cm}@{}}
\hline
Parameter & Symbol & Value & Source \\ \hline
Transmission rate to susceptible for low risk infiectious individuals (/h) & $\beta_{Low}$ & 0.005  & \cite{picault_modelling_2022} \\
 & & & \\
 Transmission rate to susceptible for medium risk infectious individuals (/h) & $\beta_{Medium}$ &$1.5 *\beta_{Low}$ & \cite{picault_modelling_2022} \\
 & & & \\
  Transmission rate to susceptible for high risk infectious individuals (/h) & $\beta_{High}$ &$1.5 *\beta_{Medium}$ & \cite{picault_modelling_2022} \\
 & & & \\
Spontaneous shedding probability & $p_E$ & 0.14, 0.42 and 0.68 for low, medium and high risk levels respectively & \cite{frank_colonization_1986} \\
 & & & \\
Initial carrier proportion & $E_0/N_0$ & 0.26, 0.48 and 0.72 for low, medium and high risk levels respectively & \cite{timsit_transmission_2013} \\
 & & & \\
Mild clinical sign duration & $\tau_M$ & 2-8 (med=6) days & \cite{grissett_structured_2015} \\
Infectious period duration & $\tau_I$ & 6-10 (med=8) days & \cite{thomas_insights_2019} \\
Asymptomatic period duration & $\tau_E$ & 1 day & \cite{grissett_structured_2015} \\
Probability of successful treatment & $p_T$ & 0.71 & \cite{dedonder_review_2015} \\
Probability of severe forms & $p_C$ & 0.65 & \cite{timsit_transmission_2013} \\ \hline
Contact rate between pens & $c$ & 0.01 & Assumed \\ \hline
\end{tabular}
\caption{Parameter values used for \textit{M. haemolytica}}
\label{table_mh}
\end{table}

\subsection*{Force of infection}
Our model enabled pathogen transmission in a building containing several pens, i.e., groups of animals sharing a same closed indoor space. We assumed that an animal in a given pen was mainly exposed to  infectious animals within its pen while also being exposed to the animals in the other pens in the building, albeit to a lesser extent.\\
The force of infection $\Phi_i$ for pen $i$ was frequency-dependent, taking into account the multi-pen nature of our model by accounting for the force of infection within the pen and between pens. Let $\mathbb{B}$ be the set of pens of $n$ individuals each raised simultaneously in a building. In each pen $i$, the intra pen force of infection was the sum of the individual contribution $\beta_\rho$ of each infectious set of individual of risk level $\rho$ ($I_{\rho,i}$). The contribution increased with the individual risk level $\rho$. Mathematically, this translates to equation (1).
 In a given pen $i$, the susceptible individuals experienced an intra pen force of infection with the addition of the contributions of the other pens of set $\mathbb{B}$ deprived of i ($\mathbb{B}\backslash \lbrace i \rbrace$) multiplied by a contact rate between pens $c$ defining how isolated the pens were from each other. In our simulations, we set $c=0.01$. This choice was motivated by the analysis of the effect of this parameter in the previous model (\cite{sorin-dupont_modeling_2023}). This value implied a low contact rate between pens, which was compliant to the low airborne transmission of \textit{M.haemolytica} (\cite{timsit_transmission_2013}). A susceptibility factor $\sigma_\rho$ increasing with the individual risk level $\rho$ multiplied this total. Mathematically, this translates to equation (2).  The force of infection $\Phi_i$ was converted to a time-dependent probability $p_{\Phi_i}$ using the equation (3), with $\delta t$ being the current time step.
 
 	\begin{equation}
 	    \forall i \in \mathbb{B}, \lambda_i= \sum_{\rho \in \mathbb{P}}{\beta_{\rho}I_{\rho,i}}
 	\end{equation}
    \begin{equation}
 	    \Phi_i=\sigma_{\rho}\frac{\lambda_i+c\sum_{b\in \mathbb{B}\backslash \lbrace i \rbrace}{\lambda_{b}}}{N_i+c(N-N_i)}
 	\end{equation}
     \begin{equation}
 	    p_{\Phi_i}(\delta t)=1 -e^{-\Phi_i \delta t}
 	\end{equation}
    
with $\mathbb{P}$ the set of individual risk levels, $\beta_{\rho}$ the shedding level of individuals with risk level $\rho$, $I_{\rho,i}$ the number of infectious individuals with risk level $\rho$ in pen $i$, $N_i$ the total number of individuals in pen $i$, $N$ the total population size. This function was proposed for modeling the force of infection as it enables to represent perfectly separated pens ($c=0$) as well as a unique large pen with homogeneous contacts ($c=1$) using a single formula. \\

\subsection*{Collective treatment criteria}
In this new model version, in addition to individual treatment of detected animals, we  also considered the possibility to trigger a collective treatment at pen scale, based on three different criteria as detailed below. Once one of the criterion was reached, all animals not already under treatment in the pen received an antimicrobial dose, and were then followed individually according to the rules described above.\\

As a first criterion for triggering a collective treatment (conventional criterion), we tested the cumulative incidence in detected BRD cases in the pen, a collective treatment being performed once this criterion reached a fixed threshold. Such a threshold has been conventionally used for a long time in French fattening operations and continues to be employed in this manner, with values usually around 10\% of the pen size (\cite{mornet_p_espinasse_j_veau_1977}), although these values are not backed up by dedicated field or experimental observations.\\

The second criterion (severity criterion) consisted in triggering a collective treatment  based on a severity function. A collective treatment was triggered when this function reached a threshold, which was selected in order to have the best trade-off possible between case incidence and antimicrobial use. The full selection process is defined in SI2. The proposed function is a weighted sum of the number of animals detected as having mild or severe clinical signs in a given pen $i$. Mathematically, this translates to equation (4):
\begin{equation}
    S_i = \alpha D_{Mild,i} + (1-\alpha)D_{C,i}, \alpha \in [0,0.5]
\end{equation}

with $D_{Mild,i}$ the number of animals detected as having mild clinical signs in pen $i$, and $D_C,i$ the number of animals detected as having severe clinical signs. The weight parameter $\alpha$ was smaller than 0.5 to allow an emphasis on severe cases. The process for selecting a relevant value for $\alpha$ is detailled in SI2.\\

The third criterion (slope criterion) considered to trigger a collective treatment was based on the speed of case appraisal. A collective treatment was triggered when this function reached a threshold, which selection answered to the same logic as the severity criterion. The slope at time $t$ in pen $i$ represented the ratio of the cumulative incidence of detected BRD cases in pen $i$ divided by the time since the beginning of the fattening. To avoid dividing by 0, $P_i(t)$ is only defined for $t>t_{0}$. This translates to equation (5):
	\begin{equation}
	    P_i (t)= \frac{cumulateD_i}{t-t_{0}}
	\end{equation}
where $cumulateD_i$ is the cumulative number of individuals detected as affected in pen $i$, $t_{0,i}$ being the starting time of the simulation, and $t$ the current time.
In all these cases, a collective treatment can be performed only once per pen over the simulation time.

\subsection*{Model implementation}

To code the model, we used the open-source \hyperlink{https://sourcesup.renater.fr/www/emulsion-public/pages/Information.html}{EMULSION framework} (\cite{picault_emulsion_2019}), which defines models in human-readable structured text files then processed by a simulation engine coded in Python. Such a modeling framework facilitates model co-development and revisions by modelers, clinicians, and epidemiologists all together. EMULSION also supports individual-based stochastic modeling, enabling simulations of small groups characterized by a high variability in biological processes. The framework allows a detailed breakdown of the modeled events and processes. Processes are articulated as finite state machines, a formalism widely employed in computer science to depict states and transitions. 

\subsection*{Scenarios and outputs}

We investigated sixteen scenarios of BRD treatment in a building of 200 young beef animals, distributed in 10 pens of 20 animals each, during 40 days, as BRD typically occurs in the first few weeks of fattening (\cite{timsit_transmission_2013,assie_exposure_2009}). These scenarios were defined by one of the four treatment strategies, applied in 2 possible farm-scale proportions of individual risk status of developing BRD and two ways of assigning animals to pens. The four treatment strategies were: (1) individual treatment administered upon detection of clinical signs (Individual); (2) collective treatment triggered when the pen-scale cumulative incidence of BRD cases reached the standard threshold, in addition to individual treatment (Conventional); (3) collective treatment triggered when the weighted severity reached the defined threshold, in addition to individual treatment (Severity); and (4) collective treatment triggered when the slope function reached the defined threshold, in addition to individual treatment (Slope). The two possible individual risk level distributions in the building were: (1) 30\%, 40\%, and 30\% of low, medium, and high risk individuals respectively (HR30); and (2) 10\%, 0\% and 90\% of low, medium and high risk individuals respectively (HR90). We chose these proportions in order to display contrasted situations, one being compliant with previously estimated risk proportion, and the other corresponding to a worst-case scenario (\cite{amrine_evaluation_2019}). Since having a majority of low risk individuals in the building would not feature enough cases to trigger a collective treatment, it was thus left out of this study. Finally, animals could either be  assigned into pens randomly (Random) or according to their individual risk level (Sorted).\\

Each scenario had 200 stochastic replicates. At the building scale (i.e. grouping the 10 pens), we observed the distribution and the median of the cumulative incidence of BRD cases and the average time spent with severe clinical signs per affected animal (in hours) with regards to the total AMU in the building. Moreover, we compared the mortality accross the scenarios. We also computed the proportion of treatment misuse for each scenario. This measure corresponds to the ratio of false positives over predicted positives, i.e in our case the ratio between the number of doses given to healthy animals over the total AMU. Furthermore, we observed the delay between the detection of the first BRD case and the collective treatment implementation. We also observed the delay between the infection and the administration of the first dose of antimicrobial on an individual scale. Finally, for the sorted scenario, we counted the proportion of pens of each risk level that triggered a collective treatment. For each of these outputs, all applicable scenarios were compared. As we could neither assume the normality of the distributions or the homogeneity of variances across the scenarios for most of the outputs, non-parametric tests were used for the analysis. General effects of treatment modalities were assessed with Kruskal-Wallis tests and post-hoc pairwise comparisons were done by using Wilcoxon-Mann-Whitney with Bonferroni correction to reduce the family-wise error rate. The signficance levels are indicated throughout the figures are the following: \begin{itemize}
    \item "****": p-value $<$ 0.0001
    \item  "***": p-value $<$ 0.001
    \item "**": p-value $<$ 0.01
    \item "*": p-value $<$ 0.05
    \item "ns": p-value $\ge$ 0.05
\end{itemize}

\subsection*{Sensitivity analysis}
To better understand the behavior of the model and to characterize the impact of parameter uncertainty, we carried out an exploration of the model sensitivity by performing a one-at-a-time sensitivity analysis on six scenarios with a common risk level proportion of 30, 40 and 30\% of low risk, medium risk and high risk respectively, as it was the most realistic (\cite{amrine_evaluation_2019}). We simulated our three collective treatment strategies in randomly allocated pens and in sorted pens.\\

The model was derived from a previous one on which a sensitivity analysis accounting for parameter interaction was performed (\cite{sorin-dupont_modeling_2023}). The latter did not provide evidence that parameter interactions contributed to model output variance. Therefore, our analysis was limited to a one-by-one analysis on the thresholds for our two new collective treatment criteria as well as the two key parameters of disease transmission: the transmission rate $\beta$ and the contact rate between pens $c$. The thresholds and the transmission rates were used at their nominal values $\pm$ 10 and 25\%. We used a larger interval for the contact rate, as little effect was observed in the sensitivity analysis of the previous model: we used the nominal value (0.01), multiplied or divided by 0.1, 0.5, 5 and 10. The tested values for each parameter are summarized in Table \ref{table_sa}. \\

\begin{table}[h!]
\centering
\begin{tabular}{@{}p{5cm}p{2.5cm}p{7cm}@{}}
\hline
 \textbf{Parameter} & \textbf{Nominal value} & \textbf{Values}              \\ \hline
Low risk transmission rate ($\beta_{Low}$) &     0.005         & $\lbrace 0.00375,0.0045,0.0055,0.00625\rbrace$ \\
Contact rate between pens ($c$)  &    0.01     & $\lbrace 0.001,0.005,0.05,0.1\rbrace$ 
\end{tabular}
\caption{Parameter values in the sensitivity analysis.}
\label{table_sa}
\end{table}

Each simulation had 200 stochastic replicates. We recorded the cumulative incidence and AMU. In order to evaluate the impat of a change in value of the transmission rate and the contact rate in each specific scenario, we computed the median of the simulations with the nominal values of the studied parameters. These medians were used as baselines. For each parameter change, we computed the relative variation of the output with regard to its respective baseline in order to evaluate the impact of the parameter change in value. We also compared the distributions of the outputs in the different collective treatment strategies with the tested parameter values in order to test the robustness of our previous conclusions. The distributions of the outputs were compared using Wilcoxon-Mann-Whitney tests with a Bonferroni correction, as normality could not be assumed. 
\newpage
\section*{Results }
\subsection*{Outputs}
\begin{figure}[h!]
    \centering
    \includegraphics[width=0.9\textwidth]{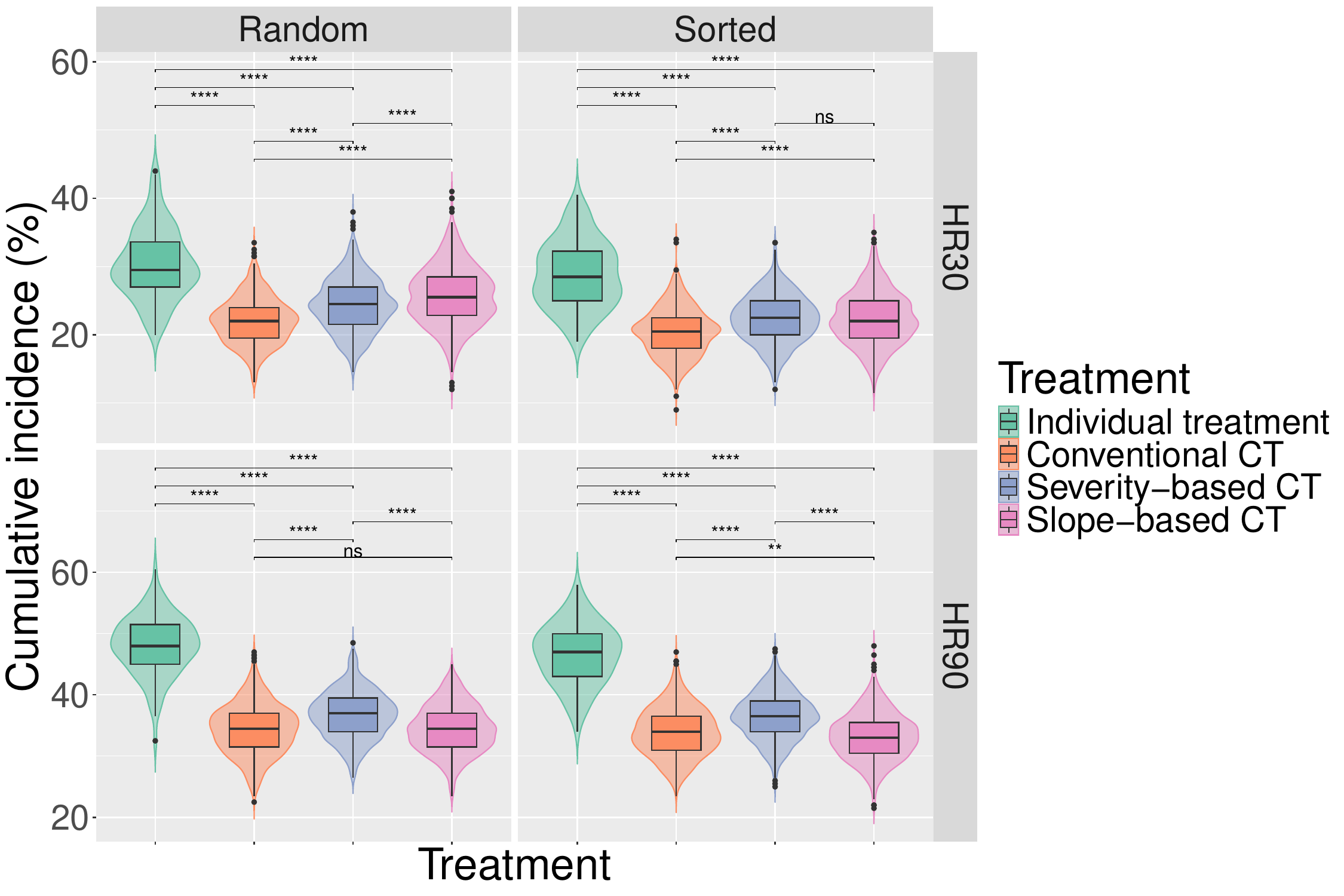}
    \caption{\textbf{Distributions of the cumulative incidence of BRD cases in the building population}. The boxplots feature the median and quartiles of the distributions. Statistical test: Wilcoxonn-Mann-Whitney test with Bonferroni's correction. First column: the animals are allocated randomly in pens. Second column: the animals are allocated in pens according to their individual risk level. Top row : 30, 40 and 30\% of low, medium and high risk animals respectively. Bottom row: 10\% of low risk animals and 90\% of high risk animals.}
    \label{incidence}
\end{figure}

Every scenario using a collective treatment yielded a lower cumulative incidence of BRD cases than their counterpart scenario using individual treatment only (Wilcoxon test, p-value < 0.05). The difference in cumulative incidence of BRD cases between using individual treatments and collective treatments was however lower in HR30 than in HR90 (Figure \ref{incidence}).\\
In HR30, the cumulative incidence was the lowest in scenarios applying the conventional collective treatment. In the scenario with a random pen allocation, the severity criterion had a lower median incidence than the slope criterion, whereas no significant difference between the two new criteria could be assessed in the scenario with sorted pens.\\
In HR90, the scenario using the slope criterion and the conventional collective treatment displayed a lower median cumulative incidence of BRD cases than the severity criterion. In the scenario with random pen allocation, no significant difference could be assessed between the conventional collective treatment criterion and the slope-based criterion whereas the latter had a significantly lower median when the pens were sorted by risk levels. Overall, sorting the pens significantly reduced the cumulative incidence in every scenario, except in HR90 for the severity criterion (Wilcoxon test, p-value <0.05).\\

\begin{figure}[h!]
    \centering
    \includegraphics[width=0.9\textwidth]{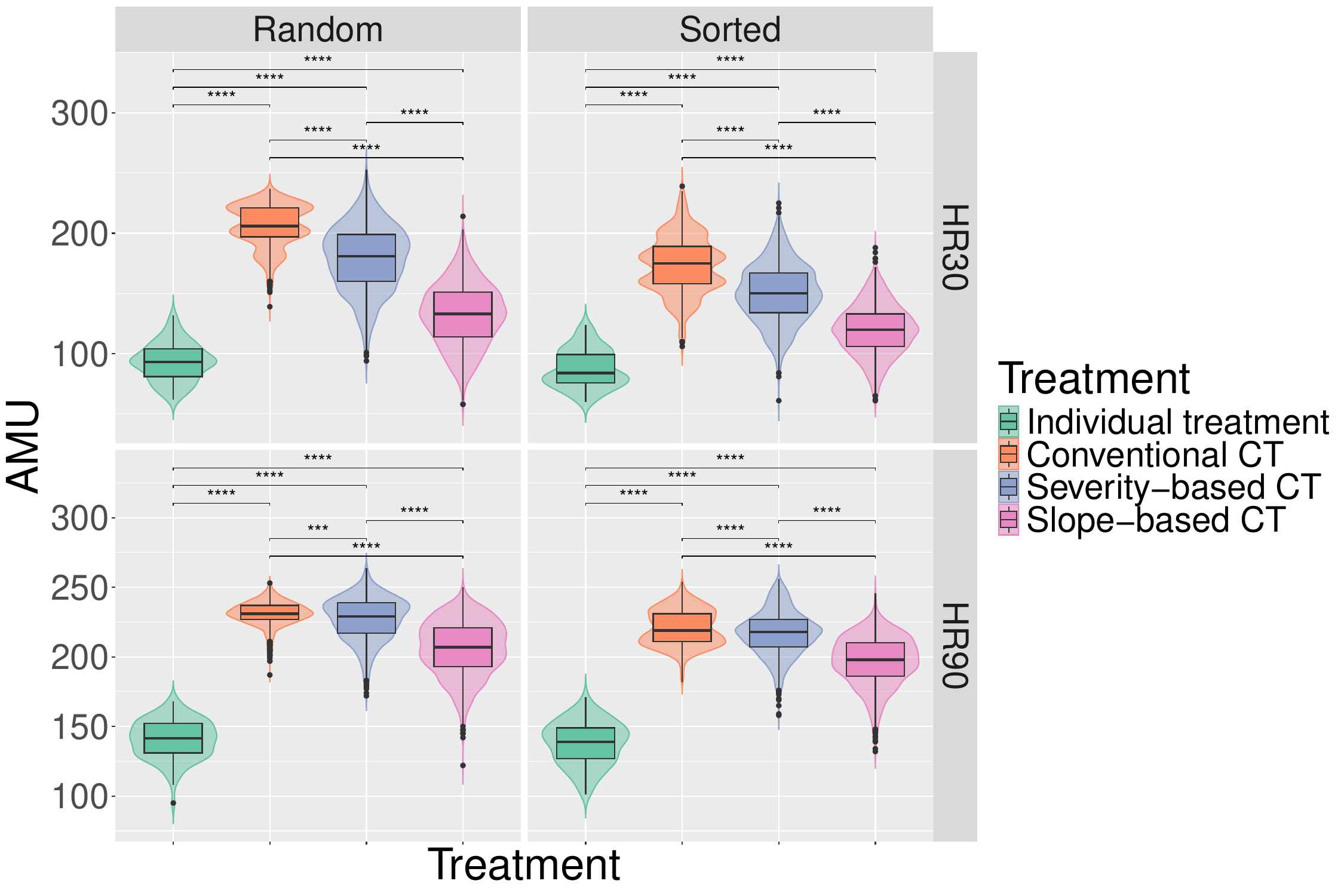}
    \caption{\textbf{Distributions of the antimicrobial use (AMU) in the building population}. The boxplots feature the median and quartiles of the distributions. Statistical test: Wilcoxonn-Mann-Whitney test with Bonferroni's correction. First column: the animals are allocated randomly in pens. Second column: the animals are allocated in pens according to their individual risk level. Top row : 30, 40 and 30\% of low, medium and high risk animals respectively. Bottom row: 10\% of low risk animals and 90\% of high risk animals}
    \label{amu}
\end{figure}

Conversely, collective treatment yielded large AMU than individual treatments (Figure \ref{amu}). In every scenario, the slope criterion yielded the lowest AMU of all the collective treatment criteria, while the conventional criterion yielded the highest. \\

\begin{figure}[h!]
    \centering
    \includegraphics[width=0.9\textwidth]{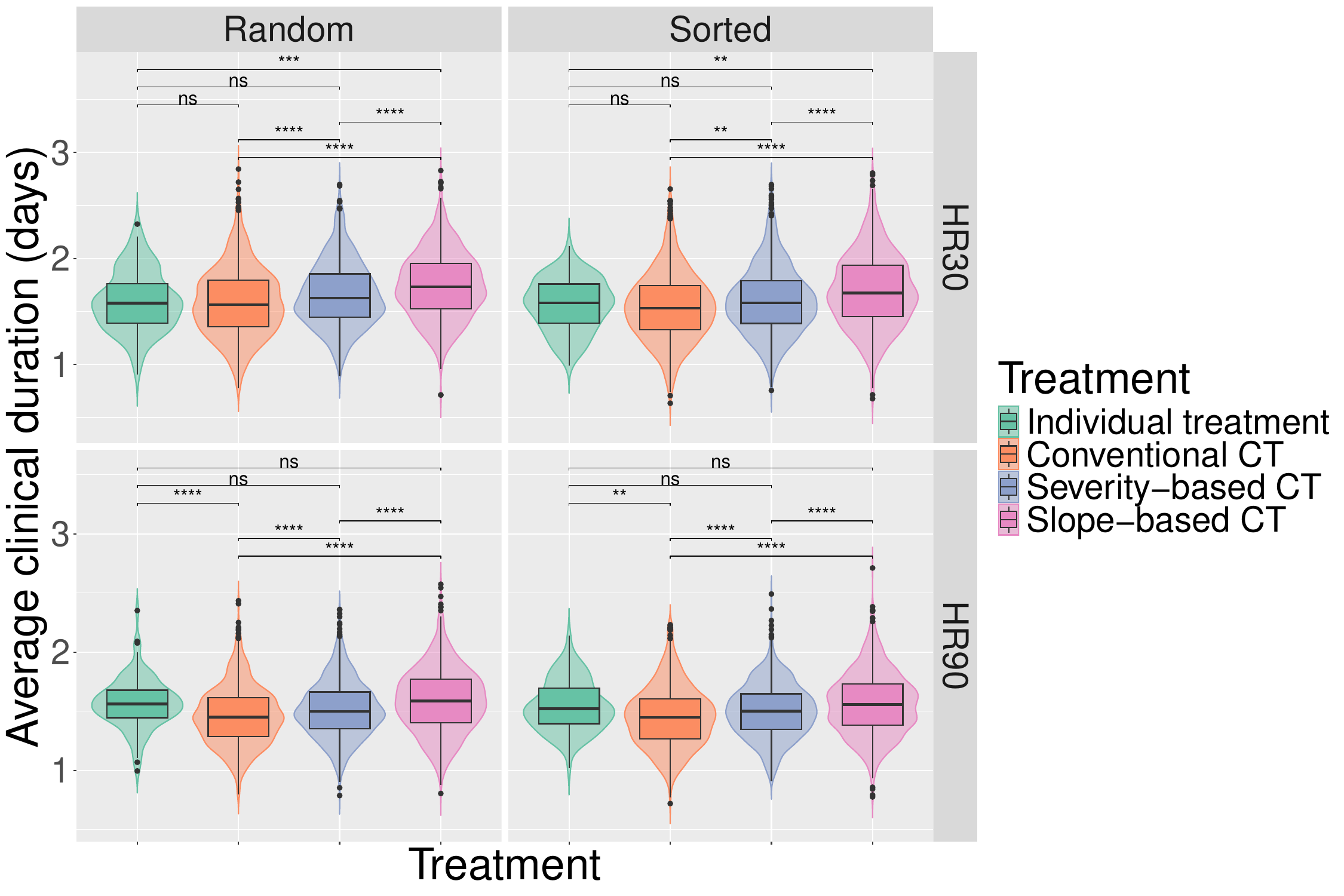}
    \caption{\textbf{Distributions of the average duration of severe clinical signs in the building population}. The boxplots feature the median and quartiles of the distributions. Statistical test: Wilcoxonn-Mann-Whitney test with Bonferroni's correction. First column: the animals are allocated randomly in pens. Second column: the animals are allocated in pens according to their individual risk level. Top row : 30, 40 and 30\% of low, medium and high risk animals respectively. Bottom row: 10\% of low risk animals and 90\% of high risk animals}
    \label{duration}
\end{figure}

The treatments appeared to have only a limited impact on the average duration of severe clinical signs (Figure \ref{duration}). Indeed, in most scenarios, we observed an important overlap of the distributions. However, the conventional criterion for collective treatment was generally associated with the lowest average duration of severe clinical signs.\\
\newpage

\begin{figure}[h!]
    \centering
    \includegraphics[width=0.9\textwidth]{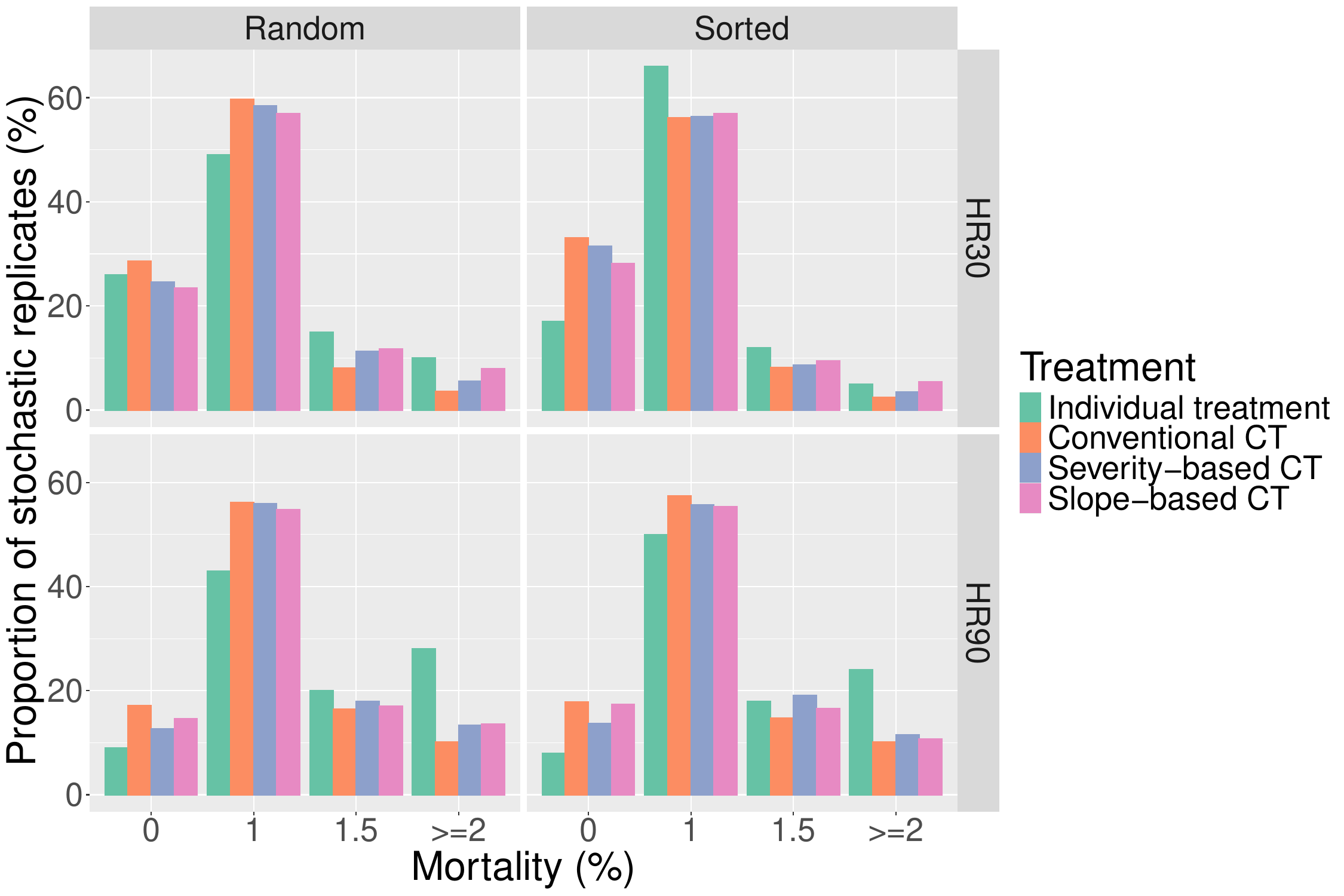}
    \caption{\textbf{Barplots of the mortality in the building population}. First column: the animals are allocated randomly in pens. Second column: the animals are allocated in pens according to their individual risk level. Top row : 30, 40 and 30\% of low, medium and high risk animals respectively. Bottom row: 10\% of low risk animals and 90\% of high risk animals}
    \label{mortality}
\end{figure}

Collective treatments successfully reduced the occurence of high mortality in HR90 scenarios when compared to individual treatments (Figure \ref{mortality}). Indeed, in random scenarios, the mortality was higher or equal to é\% in around 28\% or scenarios with individual treatment, whereas this mortality was obtained in only 10 to 13\% of replicates with collective treatments. In HR30 scenarios, the differences of in high mortality occurence are less marked, as most of the replicates yielded a mortality lower or equal to 1\%, regardless of the applied treatment strategy.\\

\begin{figure}[h!]
    \centering
    \includegraphics[width=0.9\textwidth]{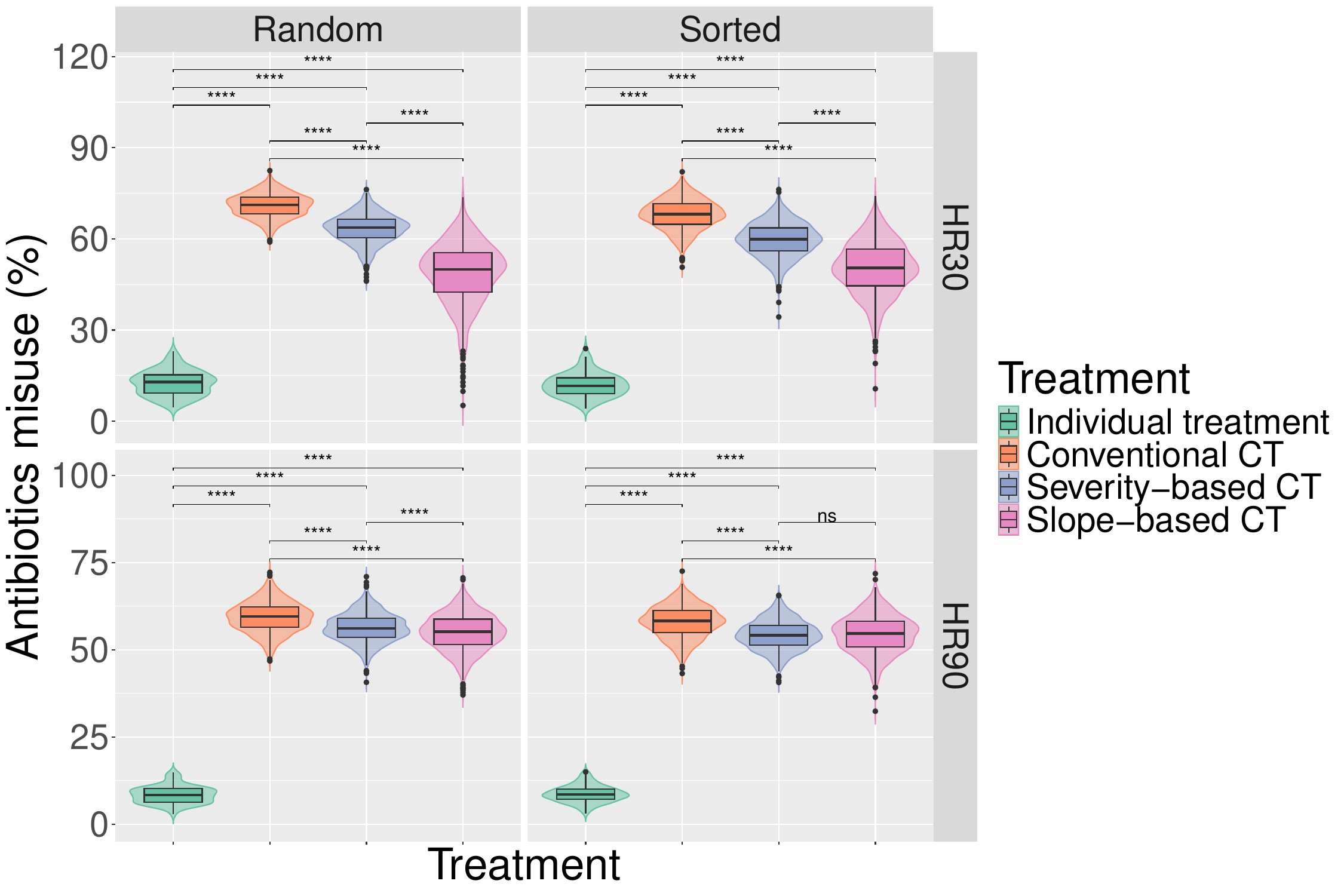}
    \caption{\textbf{Distributions of the proportion of treatment misuse}. This proportion is the ratio of doses given to healthy individuals on the total AMU. The boxplots feature the median and quartiles of the distributions. Statistical test: Wilcoxonn-Mann-Whitney test with Bonferroni's correction. First column: the animals are allocated randomly in pens. Second column: the animals are allocated in pens according to their individual risk level. Top row : 30, 40 and 30\% of low, medium and high risk animals respectively. Bottom row: 10\% of low risk animals and 90\% of high risk animals}
    \label{fdr}
\end{figure}

The proportion of treatment misuse was very low when only individual treatments were used and much higher for all collective treatments (Kruskal-Wallis test, p-value <0.05) (Figure \ref{fdr}).  In HR30, the median misuse of antimicrobial for collective treatments was between 50 and 75\%. The collective treatment with the lowest misuse appeared to be the one triggered by the slope criterion. Sorting pens appeared to have an effect on treatment misuse when collective treatments were used (Wilcoxon test, p-value <0.05). In HR90 scenarios, the conventional collective treatment also had a higher misuse than the two new criteria. In scenarios with a random pen allocation, the slope criterion was associated with the lowest misuse whereas in scenarios with sorted pens, the two new criteria were not significantly different.\\
\newpage

\begin{figure}[h!]
    \centering
    \includegraphics[width=0.9\textwidth]{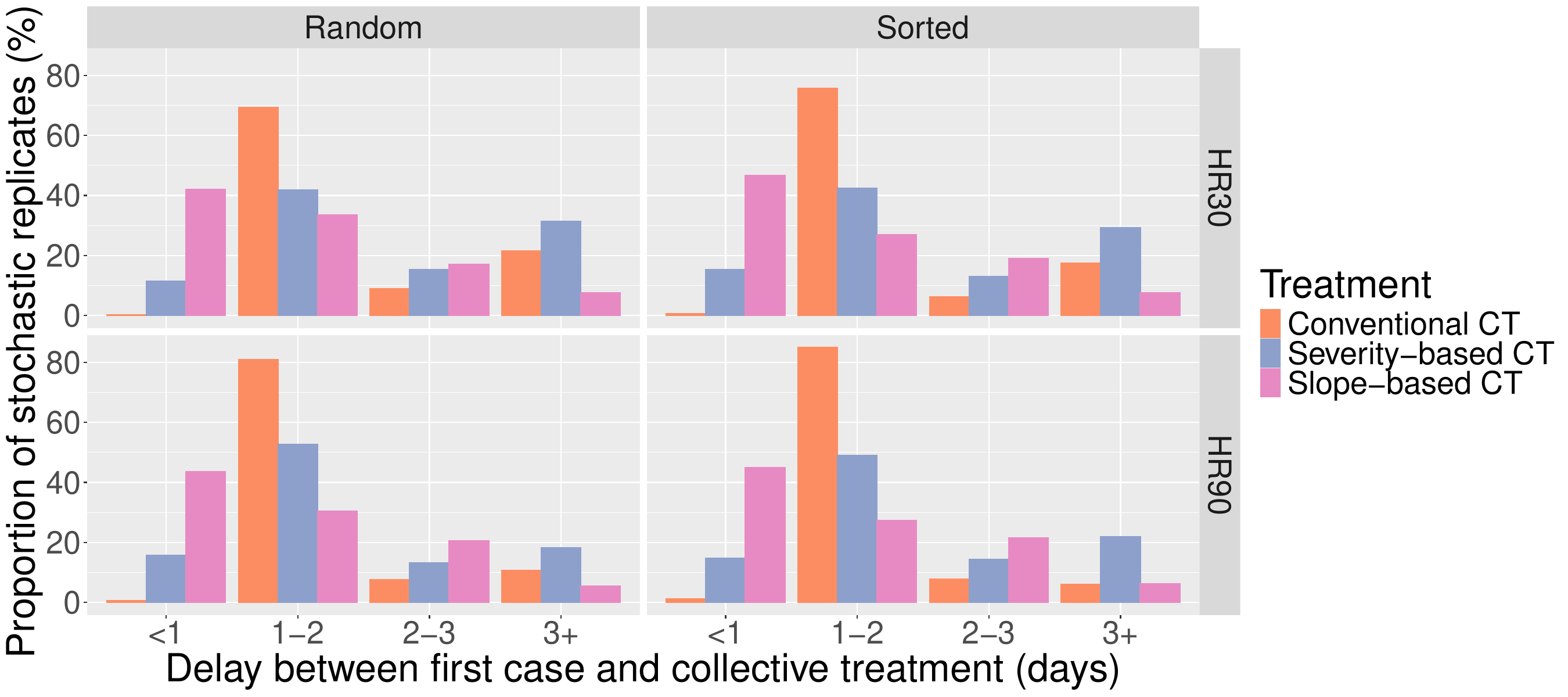}
    \caption{\textbf{Delay between the first case in a pen and the collective treatment (in days).} First column: the animals are allocated randomly in pens. Second column: the animals are allocated in pens according to their individual risk level. Top row: 30, 40 and 30\% of low, medium and high risk animals respectively. Bottom row:  90\% of high risk animals. The effect of the treatment modality on the delay is significant according to a $\chi^2$ test.}
    \label{delay}
\end{figure}

The criterion for triggering a collective treatment affected the delay between the detection of the first BRD case and the collective treatment (Figure \ref{delay}). Indeed, when collective treatments where triggered with the slope criterion, the delay between the first case and the collective treatment was shorter than a day in the majority of the stochastic replicates, whereas this delay was more often between one and two days for the conventional criterion. Collective treatment triggered by the severity criterion were triggered in less than a day in  between 26\% and 37\% of the replicates with collective treatment. In around 43\% of the cases, it was triggered between one and two days after the first case in the pen.\\

\begin{figure}[h!]
    \centering
    \includegraphics[width=0.9\textwidth]{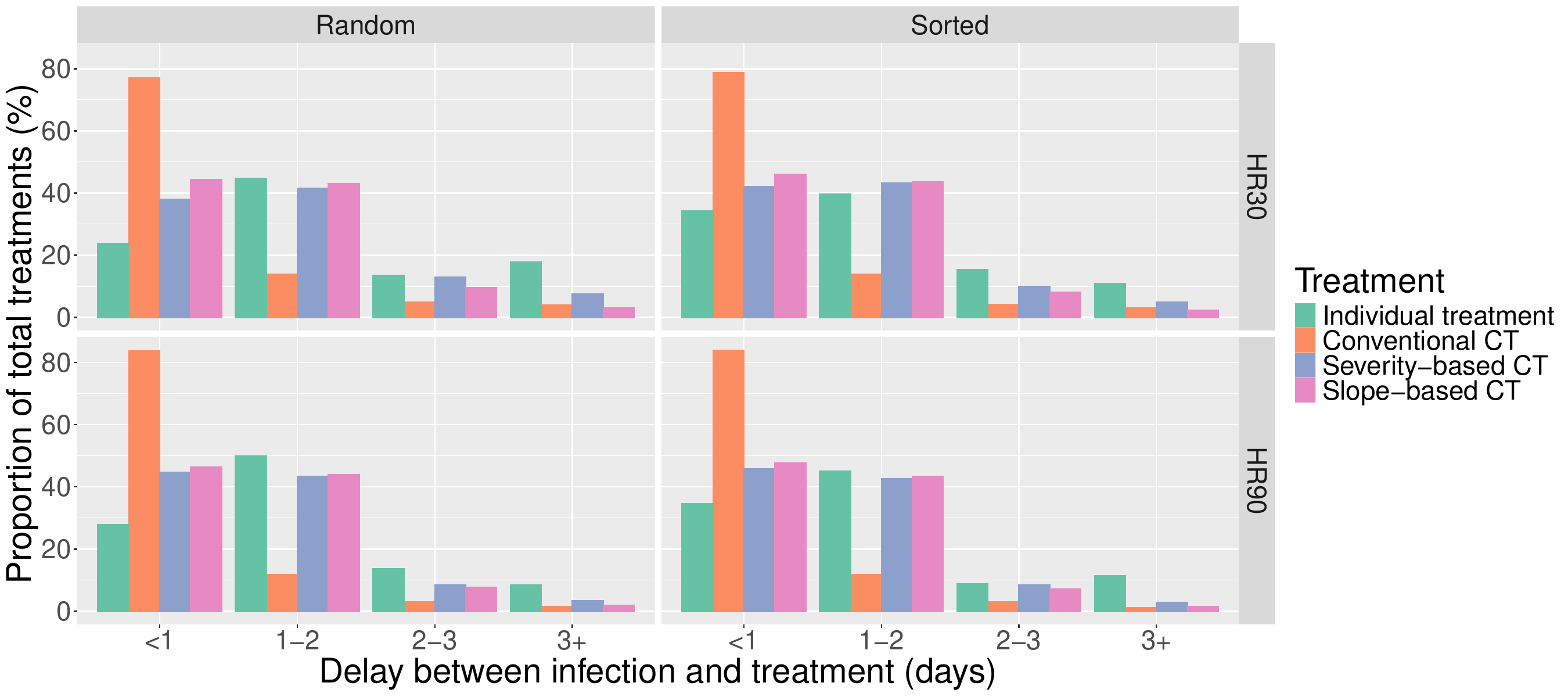}
    \caption{\textbf{Delay between infection and treatment (in days).} First column: the animals are allocated randomly in pens. Second column: the animals are allocated in pens according to their individual risk level. Top row: the population is composed of 30, 40 and 30\% of low, medium and high risk animals respectively. Bottom row: the population is composed of 10\% of low risk animals and 90\% of high risk animals.}
    \label{delay_indiv}
\end{figure}

On an individual level, collective treatments shortened the delay between the infection and the treatment (Figure\ref{delay_indiv}). The delay exceeded 2 days in less than 15\% of the replicates in scenarios with collective treatments, while delays up to 9 days were observed in scenarios with individual treatments. Among collective treatment criteria, the conventional criterion yielded the biggest proportions of delays shorter than a day (up to 84\%  in HR90 scenarios), while the other two criteria yielded similar results.\\

\pagebreak

Pens containing animals of the same individual risk level were grouped together across the simulations to observe proportions of triggered collective treatment in each specific group (Table \ref{proportions}). A collective treatment was triggered in most of the pens composed of animals with high risk of respiratory disease, irrespective of the criterion. The conventional criterion had the highest proportion of triggered collective treatments in pens composed of medium risk individuals (83.9\%), while the severity criterion triggered a collective treatment in 58.1\% of these pens and the slope criterion only triggered a collective treatment in 33.8\% of them.  In pens composed of low risk individuals, collective treatments triggered by the conventional criterion happened in 35.9\% of the pens, while collective treatments based on the severity and slope criteria only happened in 12.6 and 5.5\% of the pens respectively.   \\

\begin{table}[h!]
\begin{tabular}{@{}llll@{}}
\toprule
\multirow{2}{*}{Treatment strategy} & \multicolumn{3}{l}{Proportion of pens having triggered a collective treatment (\%)} \\ \cmidrule(l){2-4} 
                  & Low-risk pens & Medium-risk pens & High-risk pens \\ \midrule
Conventional CT   & 35.9          & 83.9             & 99.7           \\
Slope-based CT    & 5.5           & 33.8             & 85.6           \\
Severity-based CT & 12.6          & 58.1             & 91.8           \\ \bottomrule
\end{tabular}
\caption{Proportion of triggered collective treatments in sorted pens.}
\label{proportions}
\end{table}

\subsection*{Sensitivity analysis}

\begin{figure}[h!]
    \centering
    \includegraphics[width=\textwidth]{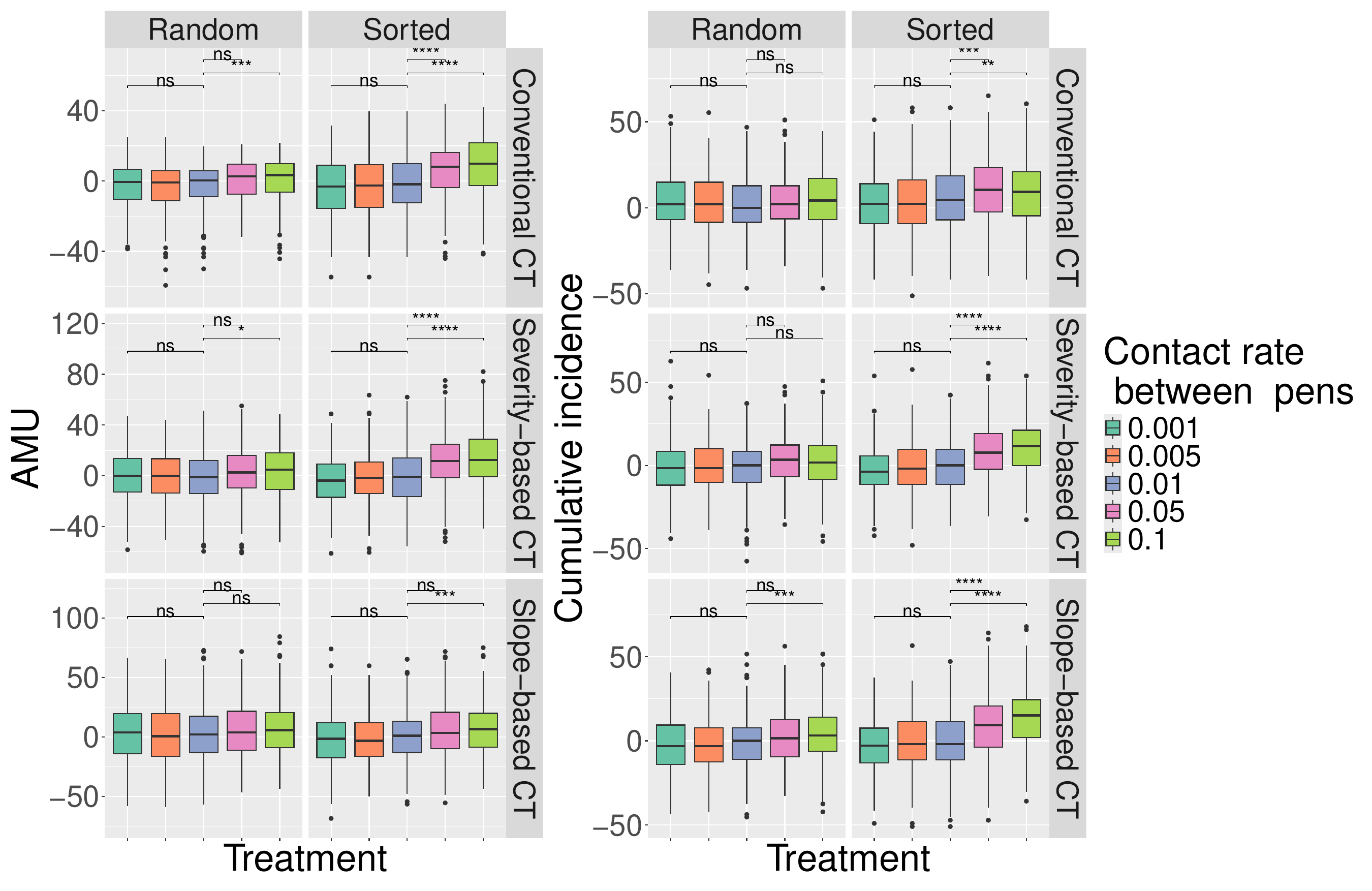}
    \caption{\textbf{Analysis of the sensitivity of the outputs to between-pens contact rate values.} Top panel: Cumulative incidence variation relative to the median of the baseline scenario -- $c=0.01$ -- (in \%). Bottom panel: AMU variation relative to the median of the baseline scenario (in \%). First row: the animals are allocated randomly in pens. Second row: the animals are allocated in pens according to their individual risk level. CT: Collective treatment. Statistical test: Wilcoxonn-Mann-Whitney test with Bonferroni's correction.}
    \label{sa_relative_contact}
\end{figure}

The between-pen contact rate had no impact on the outputs when it was lower than its nominal value (Figure \ref{sa_relative_contact}). In scenarios with a random pen allocation, a contact rate equal to 0.1 was associated with significantly higher AMU and cumulative incidence only when the slope criterion was applied. However, the scenarios with sorted pens were more sensitive to changes in contact rate values: indeed, contact rates higher than the nominal value were associated with significantly higher AMU and cumulative incidences for every collective treatment stragies.\\

\begin{figure}[h!]
    \centering
    \includegraphics[width=\textwidth]{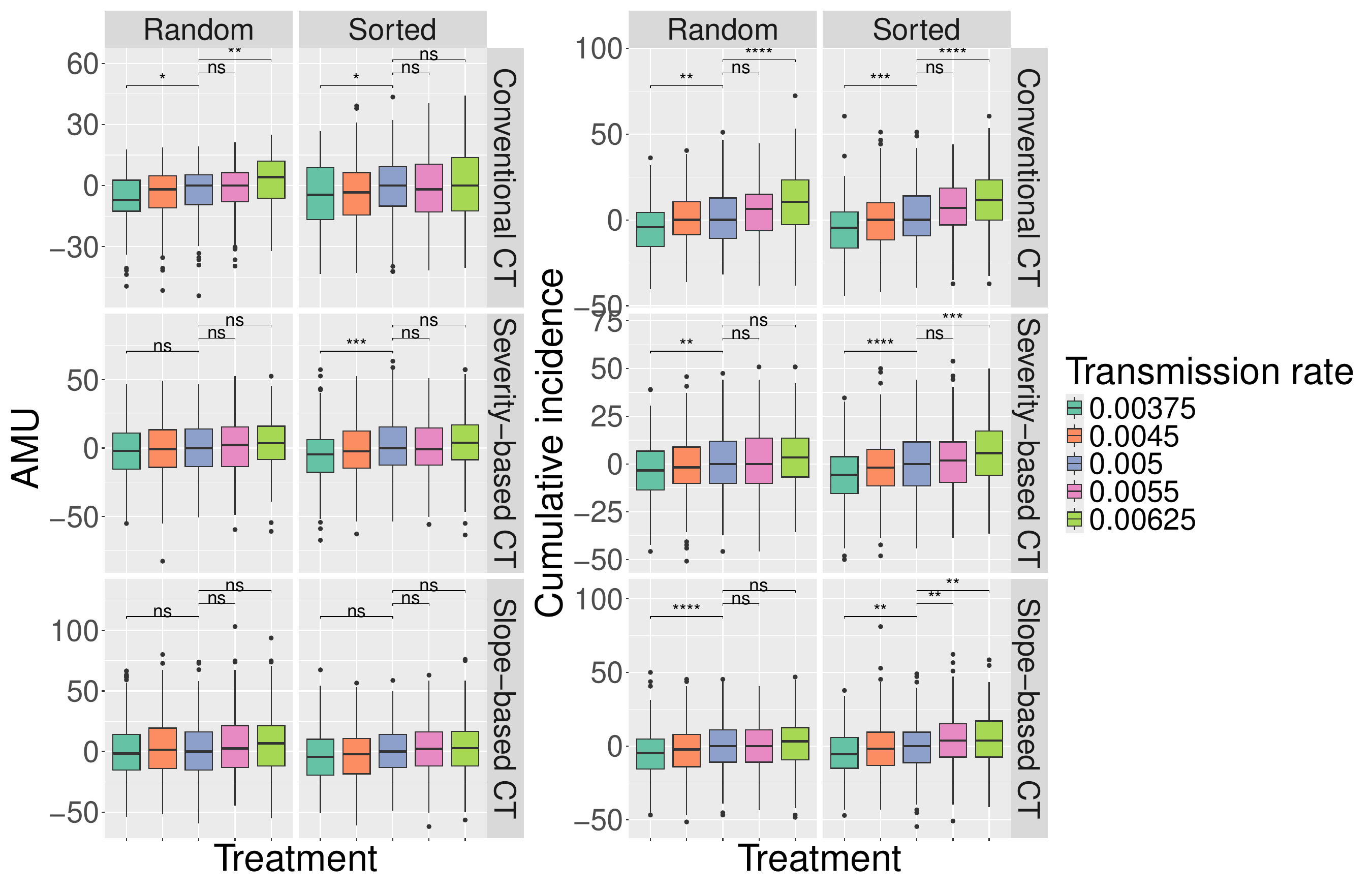}
    \caption{\textbf{Analysis of the sensitivity of the outputs to transmission rate values.} Top panel: Cumulative incidence variation relative to the median of the baseline scenario -- $\beta=0.005$ -- (in \%). Bottom panel: AMU variation relative to the median of the baseline scenario (in \%). First row: the animals are allocated randomly in pens. Second row: the animals are allocated in pens according to their individual risk level. CT: Collective treatment. Statistical test: Wilcoxonn-Mann-Whitney test with Bonferroni's correction.}
    \label{sa_relative_rate}
\end{figure}

The transmission rate variations yielded a stronger impact on the cumulative incidence than on the AMU (Figure \ref{sa_relative_rate}). Indeed, the AMU was significantly lower when the transmission rate was equal to 0.00375, with no other clear differences for the other transmission rate values. In particular, the AMU for the slope criterion remained unchanged for every transmission rate value. The cumulative incidence was similarly affected by lower transmission rate values. Additionnaly, it was also sensitive to transmission rates higher than the nominal value, especially in scenarios with sorted pens. Indeed, in scenarios with sorted pens, the cumulative incidence was significantly higher than the baseline scenario when the transmission rate was equal to 0.00625 for every collective treatment criterion.

\begin{figure}[h!]
    \centering
    \includegraphics[width=\textwidth]{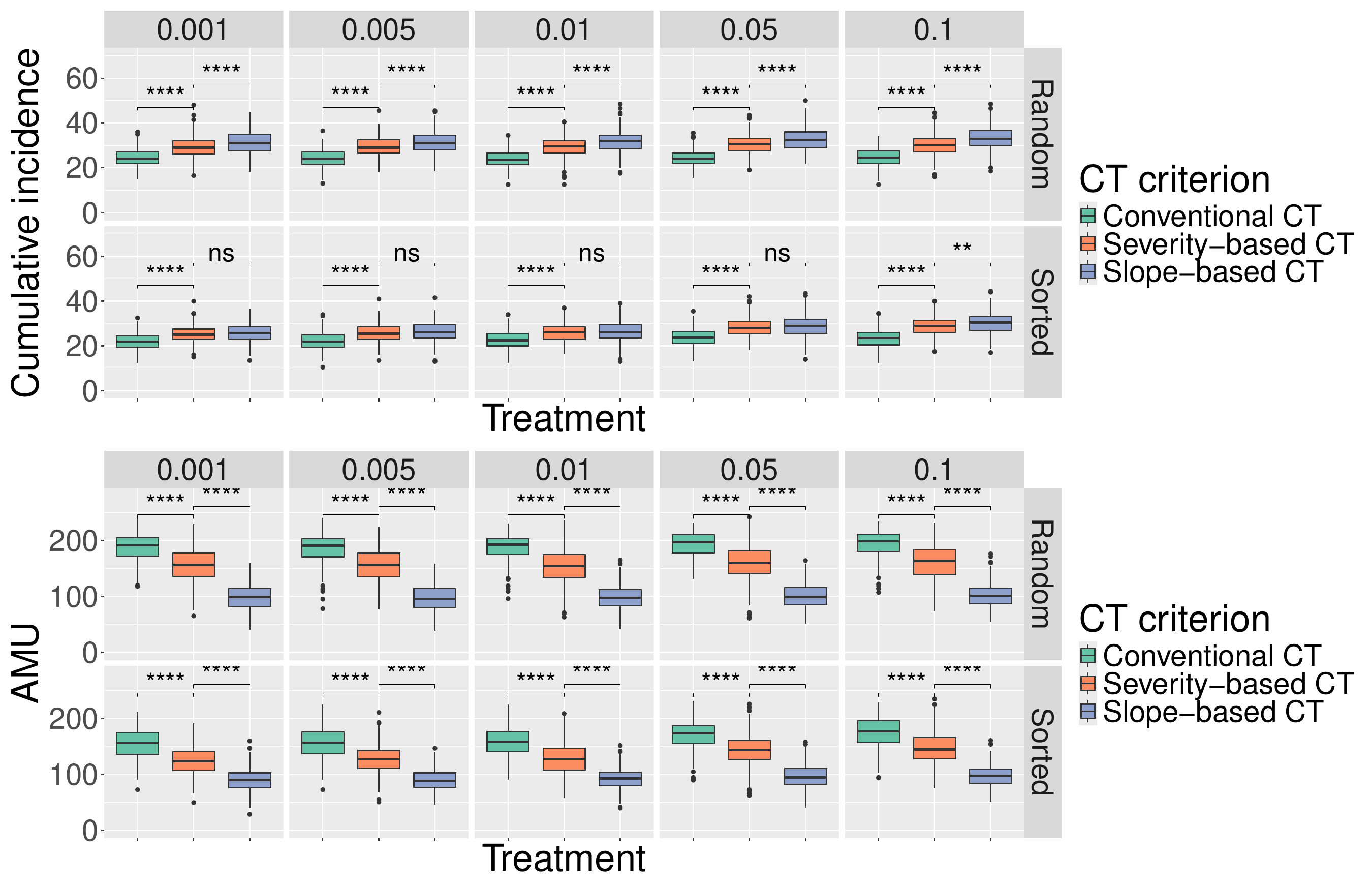}
    \caption{\textbf{Impact of the collective treatment strategies on the outputs with the tested  between-pen contact rate values.} Top panel: Cumulative incidence. Bottom panel: AMU. First row: the animals are allocated randomly in pens. Second row: the animals are allocated in pens according to their individual risk level. Each column represents the tested value for the contact rate. Statistical test: Wilcoxonn-Mann-Whitney test with Bonferroni's correction..}
    \label{sa_contact}
\end{figure}

Changes in contact rates between pens had little to no impact on the ranking of the collective treatment criterion (Figure \ref{sa_contact}). Indeed, For every contact rate value, the slope criterion was always associated with the highest cumulative incidence in scenarios with random pens and was associated with the lowest AMU.\\
\newpage

Similar results were obtained when comparing the collective treatment criteria under several transmission rates (Figure \ref{sa_rate}). Indeed, the slope criterion was still associated with the lowest AMU and the highest incidence in random pen allocation scenarios.\\

\newpage

\section*{Discussion}

This study aimed at modeling the impact of triggering collective antimicrobial treatments according to three different criteria on the spread of \textit{M. haemolytica}  in multi-pens fattening systems while accounting for the multi-pen structure, individual risk of being affected by BRD and farming practices in terms of pens composition.  We computed the cumulative incidence, the average duration of severe clinical signs and the mortality. We balanced these outputs  against the AMU and treatment precision for each scenario. We also computed the delay between the beginning of an episode and the beginning of the treatment and calculated the proportion of pens triggering collective treatments based on each criterion. We proposed that a novel criterion, based on the speed of the spread rather than on a proportion of detected individuals, could yield a better trade-off between BRD case incidence and AMU than the conventional way of implementing collective treatment. This result could have interesting operational implications on the decision-making regarding collective treatment in fattening farms.\\

We showed that collective treatment could efficiently reduce the incidence of bacterial respiratory disease in population with high individual risk of BRD.  The positive impact on animal health and welfare balanced the supplementary cost in antimicrobial doses. This compromise in favour of reducing BRD impact was particularly relevant and valuable for high-risk individuals. This finding is consistent with a previous simulation study which implemented collective treatments on an ‘average pathogen’ with less strict thresholds (\cite{picault_modelling_2022}). In American systems, medication of at-risk cattle is frequent at feedlot entry after their idenfication and the implementation of preconditionning programs (\cite{gonzalez-martin_reducing_2011}). These programs has been studied in Europe, but the results remained inconclusive regarding their positive impact on animal health and welfare (\cite{vanbergue_comparison_2024}). The study however concluded that the cattle sector should focus its effort on improving husbandry practices. This suggestion aligns with the use of a selective criterion for the implementation of curative collective treatments. \\

Our results showcased a positive impact of collective treatments on disease related mortality in the herd in high risk scenarios. However, no criterion was identified as systematically better than the others in reducing the mortality. This can be due to the fact that collective treatments happened rather rapidly after an outbreak, in spite of some differences in delay. This swift reaction did not let time for animals to develop severe clinical signs and to die subsequently to them. Our results aligned with field studies showing reduced mortality in herds where collective treatment was used (\cite{oconnor_mixed_2013}).\\

The supplementary cost in antimicrobial was less desirable when the general risk was lower in the population, as the reduction in incidence and severity thanks to collective treatment was less substantial. The overuse of antimicrobial leads to antimicrobial resistance and treatment purchase and administration are costs supported by the farmer (\cite{laxminarayan_antibiotic_2013}). Therefore, there appears to be a trade-off between incidence reduction and antimicrobial use which could benefit from a characterization of the risk level of the pens. Such a characterization could guide the farmers in making the best decision regarding treatment and herd management (\cite{edwards_control_2010}).\\

Our model allowed us to identify a beneficial novel criterion for collective treatment. Indeed, instead of a threshold of detected animals in a pen, basing a collective treatment decision on the speed of the spread of the disease allowed marginal gains in disease incidence while drastically reducing the AMU and the treatment misuse when compared to the other collective treatment criteria. A field meta-analysis of randomized clinical trials showed that the risk reduction of BRD varied drastically with the criterion for collective treatment (\cite{baptiste_antimicrobial_2017}). The routine criterion is 10\% of morbidity for two consecutive days, however, there is no evidence based studies proving the relevance of this threshold (\cite{radostits_herd_2001}). Perspectives of this study include the investigation of the novel treatment strategies in broader contexts in order to translate the thresholds used into operative indicators.\\

Collective treatments based on the speed of the spread of the pathogen proved to be triggered sooner than with the other criteria. This precocity was key for the reduction of the incidence and severity of the disease, especially in high-risk populations. It also showed to target more specifically high-risk groups than the other criteria, as it reduced the proportion of low-risk pens having received collective treatments. On an individual level, the delay between the infection and the administration of the first dose of treatment was also characterized. This delay was shortened by the implementation of collective treatment strategies, particularly for the conventional criterion. These are original findings with clear practical implications, especially with regards to the need for a better knowledge by farmers of individual risk of animals developing BRD, in order to account for this information when constituting pens. However, triggering collective treatments greatly relies on the ability to detect diseased animals in the first place, therefore, increasing the sensitivity via standardized detection procedures may increase the benefit of the subsequent control measures and may modify the most convenient criteria to trigger collective treatment (\cite{kamel2024strategies}).\\

The novel slope criterion proposed by this study also outperformed the conventional criterion in terms of antimicrobial misuse, especially in moderate risk scenarios. As the animals with a high individual risk of BRD are more prone to develop and to transmit respiratory pathogens, targetting the most at risk pens is a desirable trait for BRD control measures. Our results on collective treatments occurrence in sorted pens highlight the ability of the novel slope criteria to target the most at risk pens, therefore reducing the antimicrobial use in pens with a lesser risk of an outbreak. This also translated to the lowest proportion of treatment misuse at herd level when compared to the other two collective treatment criteria. In scenarios with individual treatments only, the misuse of antimicrobial on healthy animals was even lower than for the slope criterion, as it was solely due to the false discovery of diseased animals via temperature check, as hyperthermia could be non specific to infection.\\

The conclusions brought up by our model proved to be robust to changes in transmission rates and between-pen contact rates. Furthermore, it also brought evidence that the slope criterion benefited from sorted pens, as this setting allowed the incidence and AMU to be particularly stable. These findings allowed us to prove the robustness of our results and to further emphasize the potential of the new collective treatment criterion we brought forward.\\

Our model simulated scenarios assuming the circulation of only one bacterial pathogen. BRD is essentially a multi-pathogen disease, and synergies between viruses and bacteria have been reported to change the disease dynamics and severity (\cite{gaudino_understanding_2022})
. Further investigations of the effect of treatment strategies could be conducted in a context with mixed infections.  However, quantifying the treatment response in such context is still a pending question (\cite{fulton_lung_2009}).\\

There is clear evidence that antimicrobial resistance is rising among \textit{M.haemolytica} populations (\cite{woolums_multidrug_2018,timsit_prevalence_2017}). This rise happens on time scales that are longer than the one we modeled. Therefore, we modeled the existence of resistant strains by calculating a probability of treatment success, which value was set along the span of the simulation. Future work could aim at evaluating the long term effects of collective treatment protocols on the antimicrobial trends. \\

The force of infection used in this model assumed an increased pathogen transmission within pens with an additional equal contribution of the surrounding pens. This hypothesesis relied on the fact that aerosolization of \textit{M.haemolytica} has been pointed as a contributing factor to rapid disease spread among cattle populations in close quarters (\cite{rice_mannheimia_2007}). However, this airborne transmission is not as effective as transmission via close contacts. Adaptions of the model could therefore implement a force of infection based on the farm spatial organization (\cite{ackermann2000response}).\\

Our study advocates for more transparency on the individual risk level for BRD in the cattle production system. Such a characterization could allow mindful investments for the farmers as well as more adequate control measures regarding housing and treatment. Indeed, our novel collective treatment criterion proved to be able to target at-risk pens while not applying collective treatments to lower risk pens.\\

This model provides a solid basis for testing these new criteria for triggering collective treatments vs. keeping only individual ones on other known respiratory pathogens or a combination thereof, in order to identify potential new operational recommendations on treatment strategies. Future work could also test the inclusion of automated tools for decision making regarding treatment.

\section*{Acknowledgements}

 We thank our colleagues from the BIOEPAR Dynamo team for their comments and advice on this work. We are most grateful to the INRAE MIGALE bioinformatics facility  (\cite{noauthor_migale_nodate}), for the use of their computing resources.

\section*{Fundings}

This work has received funding from the European Union's Horizon 2020 research and innovation program under grant agreement No. 101000494 (DECIDE) and from the French region Pays de Loire. This work was also supported by the French region Pays de la Loire (PULSAR grant).

\section*{Conflict of interest disclosure}

Pauline Ezanno is recommender of PCI Animal Science.

\section*{Data, script and code availability}
The complete code for the model model can be found on this git repository: \url{https://forgemia.inra.fr/dynamo/brd-models/brd-public/-/tree/PCI_animal_science2024}. The version of the model used for this work is tagged under \texttt{pci\_animal\_science\_2024}. 

\newpage
\appendix
\newpage

\newpage

\section{SI1: Model processes}

\subsection{Implementation and formalism of the model}
The model is a stochastic individual based model. In the model, five processes drive the states of individuals: infection, hyperthermia, clinical signs, detection and treatment. They are represented by a formalism broadly used in computer science, finite state machines, which is close to flow diagrams used by epidemiologists, with a higher expressiveness. They represent finite state an individual can find itself in.  Their textual description in the YAML format in the model file is easy to read and can be automatically converted in a graphical representation by EMULSION. The model can be found on a git repository, accessible by following this link: \url{https://forgemia.inra.fr/spicault/brd-withinherd-emulsion.git}. The version of the model used for this work is tagged under \texttt{pci\_animal\_science\_2024}. \\

To help understand the state machine diagrams, we present how the diagram summarising the infection process relates to the state machine specification, extracted from the whole model. More details on how state machines differ from flow diagrams are available in the documentation: \url{https://sourcesup.renater.fr/www/emulsion-public/pages/Modelling_principles.html}.

\newpage

\section{SI2: Threshold selection for the Novel criteria for collective treatment}
\subsection{Method}
In scenarios featuring collective treatment triggered by slope and severity criteria, the collective treatment was implemented when the respective criterion went above a certain threshold. Such a threshold thus had to be defined as a preliminary step of our study.\\

\begin{figure}[h!]
    \centering
    \includegraphics[width=\textwidth]{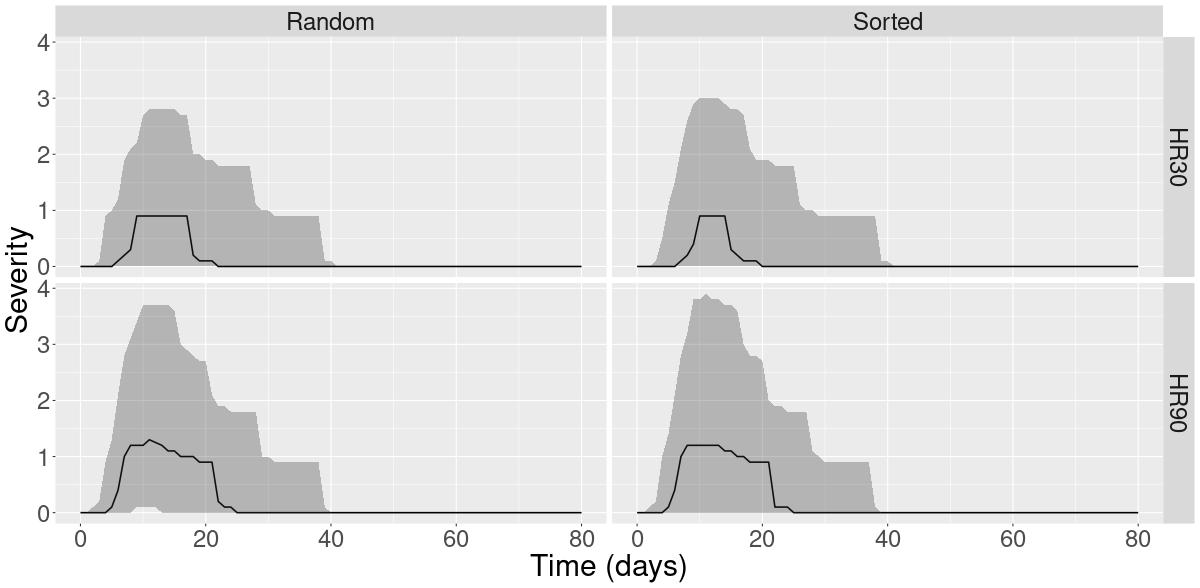}
    \caption{Median and 5-95\% quantiles of the severity criterion ($\alpha=0.4$) across scenarios with individual treatments only. The values recorded for the exploration interval were the maxima of the median and 95\% quantile in the HR30-Random scenario}
    \label{1E}
\end{figure}

\begin{figure}[h!]
    \centering
    \includegraphics[width=\textwidth]{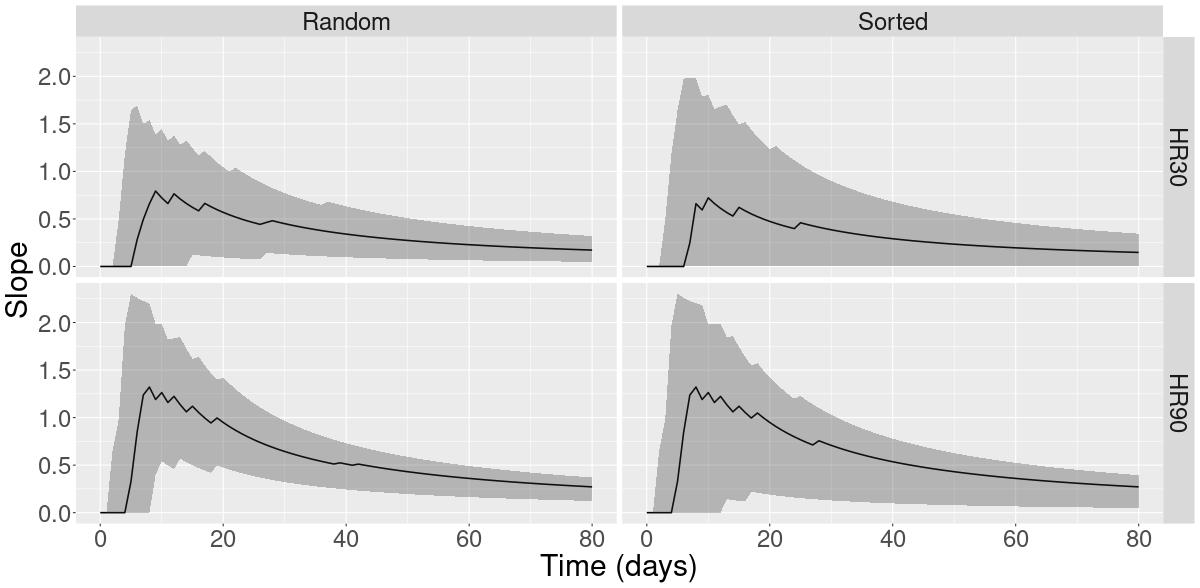}
    \caption{Median and 5-95\% quantiles of the slope criterion across scenarios with individual treatments only. The values recorded for the exploration interval were the maxima of the median and 95\% quantile in the HR30-Random scenario }
    \label{2E}
\end{figure}

In order to determine the relevant threshold for intervention, we defined a range of viable threshold values. We thus plotted both criteria on baseline scenarios (without any collective treatments). This baseline scenarios was repeated along 200 stochastic replicates in order to obtain a distribution of values for each time step. For each time step, the median value, the 5th and 95th percentiles were recorded, thus resulting in three time series for each criterion and each scenario (Figures \ref{1E},\ref{2E}). \\

The interval was extracted from the HR30 Random scenario for both criteria. The upper bound was defined as the maximum point of the 95\% quantile time series and the lower bound was defined by the maximum point of the median time series. The rationale behind this was to exclude values that would either be too low or too likely while also excluding any values that would be too high or too unlikely. Indeed, HR30 Random represents a baseline scenario and our selection allowed us to select only the higher half of the maximum values in this scenario. Thus, we obtained a credible range of $[1,3]$ for the severity criterion, discretized with intervals of 0.2. This criterion is also characterized by  parameter $\alpha$, assumed strictly lower than 0.5 to emphasize the number of individuals with severe clinical signs. This parameter was assigned a credible range of $[0.1,0.4]$ with intervals of 0.1.\\
We used the same rationale for the selection of a credible range for the slope criterion. We thus selected a range encompassing the maxium value of the median run in the HR30-Random scenario and the maxium of the 95\% quantile of this scenario. For the slope criterion, the credible range was thus $[0.75,1.7]$, discretized by intervals of 0.05. \\
We tested each threshold value in scenarios defined by 3 possible farm-scale proportions of individual risk status of developing BRD and two ways of assigning animals to pens.\begin{itemize}
    \item 90\%, 0\% and 10\% of low, medium and high risk respectively (HR10). This scenario was used as a 'best case scenario', that should not be triggering collective treatment, thus allowing to discriminate thresholds that would be too sensitive.
    \item 30\%, 40\%, and 30\% of low, medium, and high risk respectively (HR30)
    \item 10\%, 0\% and 90\% of low, medium and high risk respectively (HR90)
\end{itemize}  
Additionally, animals could be randomly assigned into pens (Random) or be assigned according to their individual risk level (Sorted).\\

For each threshold value in each scenario, we computed the median cumulative incidence and AMU in order to identify which threshold value had the best trade-off between case incidence and AMU. We determined the cutoff values of the cumulative incidence and AMU by computing the mean of both of the outputs accross the tested thresholds. We then selected the threshold value equal or inferior to the mean of both outputs.

\subsection{Results}
\subsubsection{Severity criterion}
\begin{figure}[h!]
    \centering
    \includegraphics[width=0.85\textwidth]{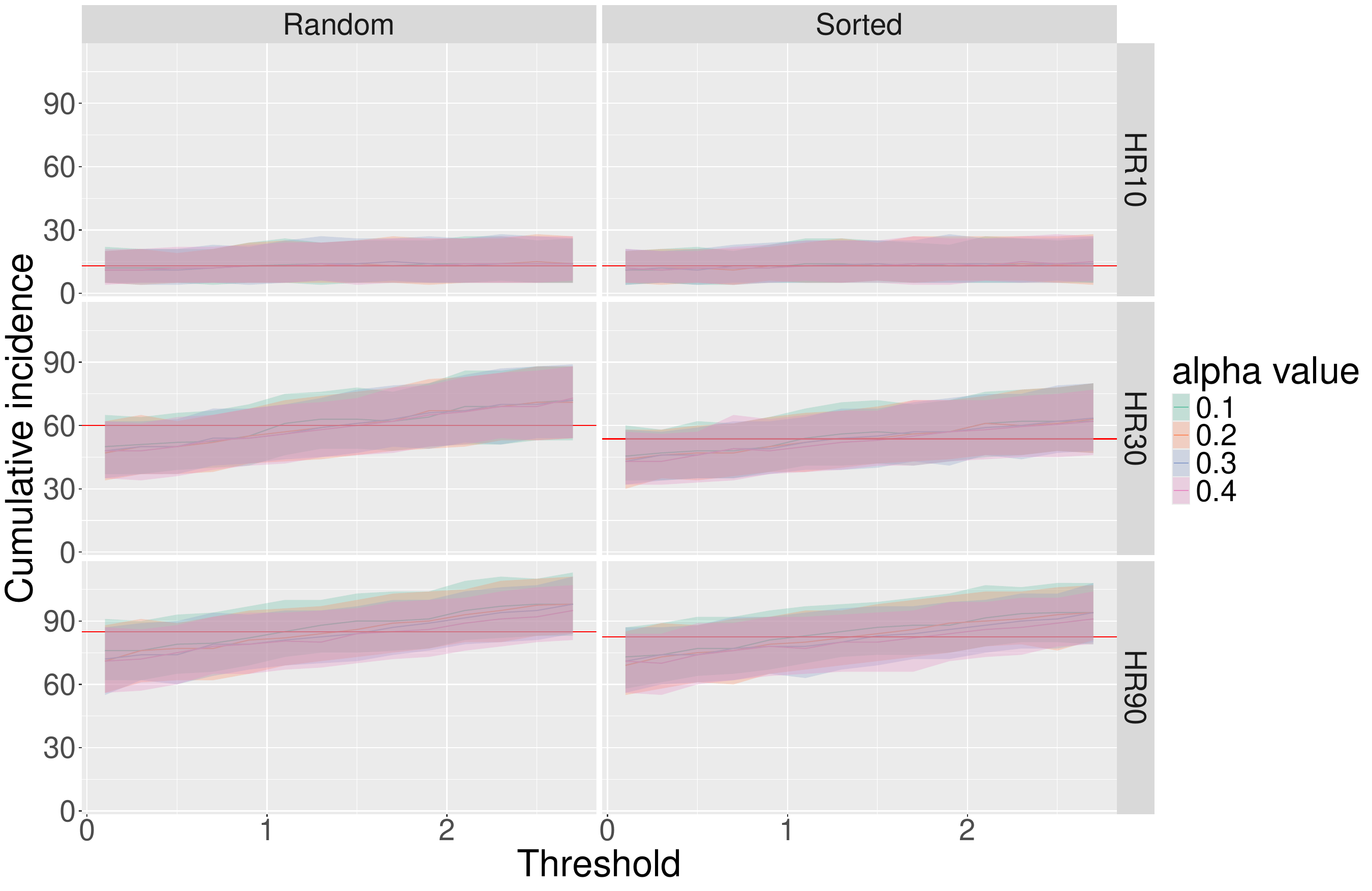}
    \caption{Median and 5-95\% quantiles of the cumulative incidence across scenarios and threshold values for the severity based criterion. The horizontal red line represents the mean value of the cumulative incidence accross the thresholds in the given scenario.}
    \label{4E}
\end{figure}

In HR10 scenarios, the cumulative incidence was roughly constant across the threshold values (Figure \ref{4E}). In HR30 and HR90 scenarios, it increased linearly. The incidence was superior to the cutoff in every scenario when the threshold was higher than 1.5.\\

\begin{figure}[h!]
    \centering
    \includegraphics[width=0.85\textwidth]{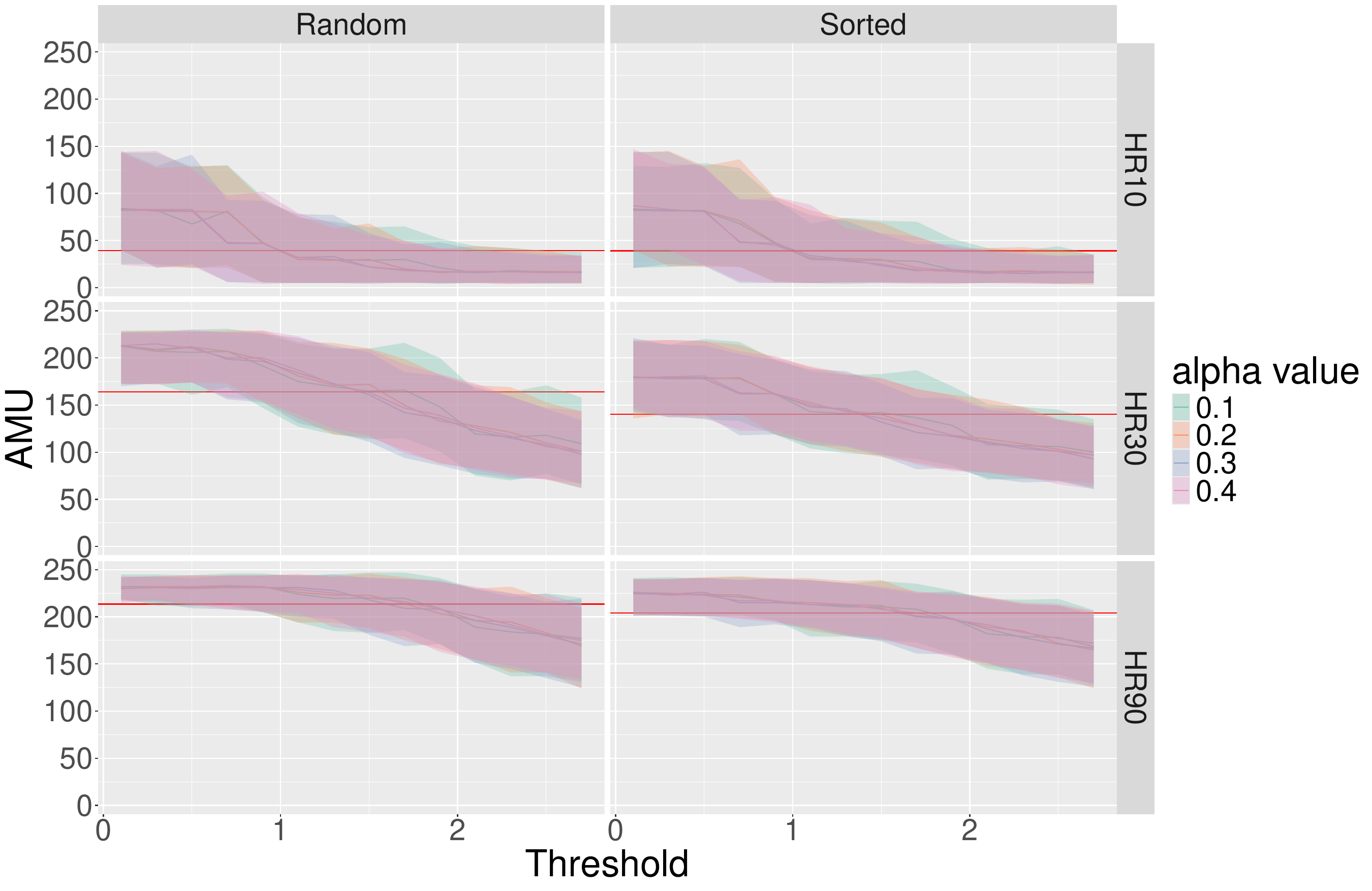}
    \caption{Median and 5-95\% quantiles of the AMU across scenarios and threshold values for the severity based criterion. The horizontal red line represents the mean value of the AMU accross the thresholds in the given scenario.}
    \label{5E}
\end{figure}

In HR10 scenarios, the AMU had a steady decrease for low threshold values and then plateaued to values lower than the cutoff (Figure \ref{5E}). Conversely, we observed a plateau of high AMU with low threshold values in HR30 and HR90 scnearios. Thresholds values lower than 1.5 were either associated with AMU higher than the low plateau in HR10 or entering the high plateau in HR30 and HR90. Therefore, we chose 1.5 as a cutoff value for the severity criterion.\\

\subsubsection{Slope criterion}
\begin{figure}[h!]
    \centering
    \includegraphics[width=0.85\textwidth]{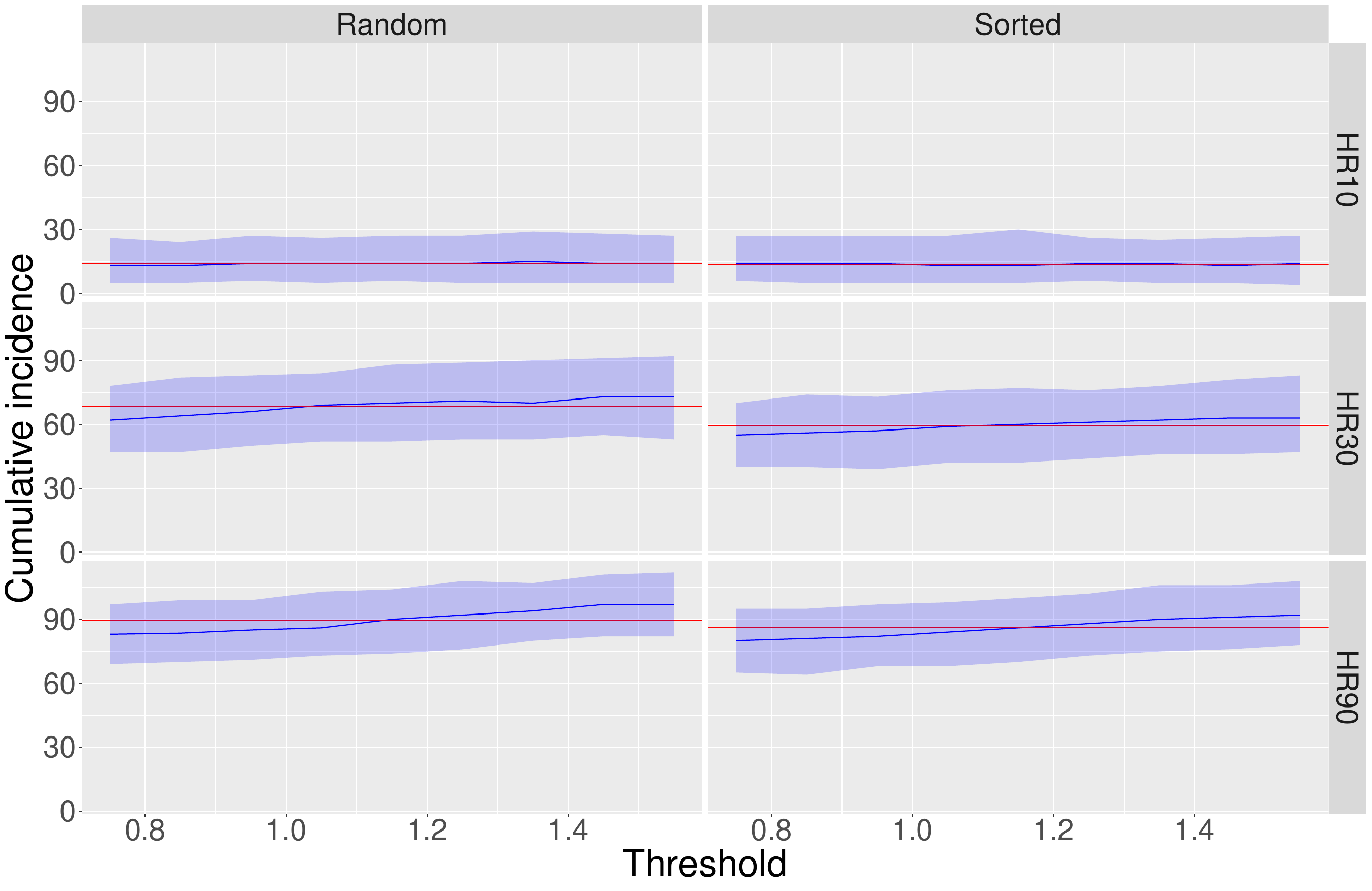}
    \caption{Median and 5-95\% quantiles of the cumulative incidence across scenarios and threshold values for the slope based criterion. The horizontal red line represents the mean value of the cumulative incidence accross the thresholds in the given scenario.}
    \label{4F}
\end{figure}

In HR10 scenarios, the cumulative incidence was also roughly constant across the threshold values (Figure \ref{4F}). In HR30 and HR90 scenarios, it increased steadily. The incidence was superior to the cutoff in every scenario when the threshold was higher than 1.15.\\

\begin{figure}[h!]
    \centering
    \includegraphics[width=0.85\textwidth]{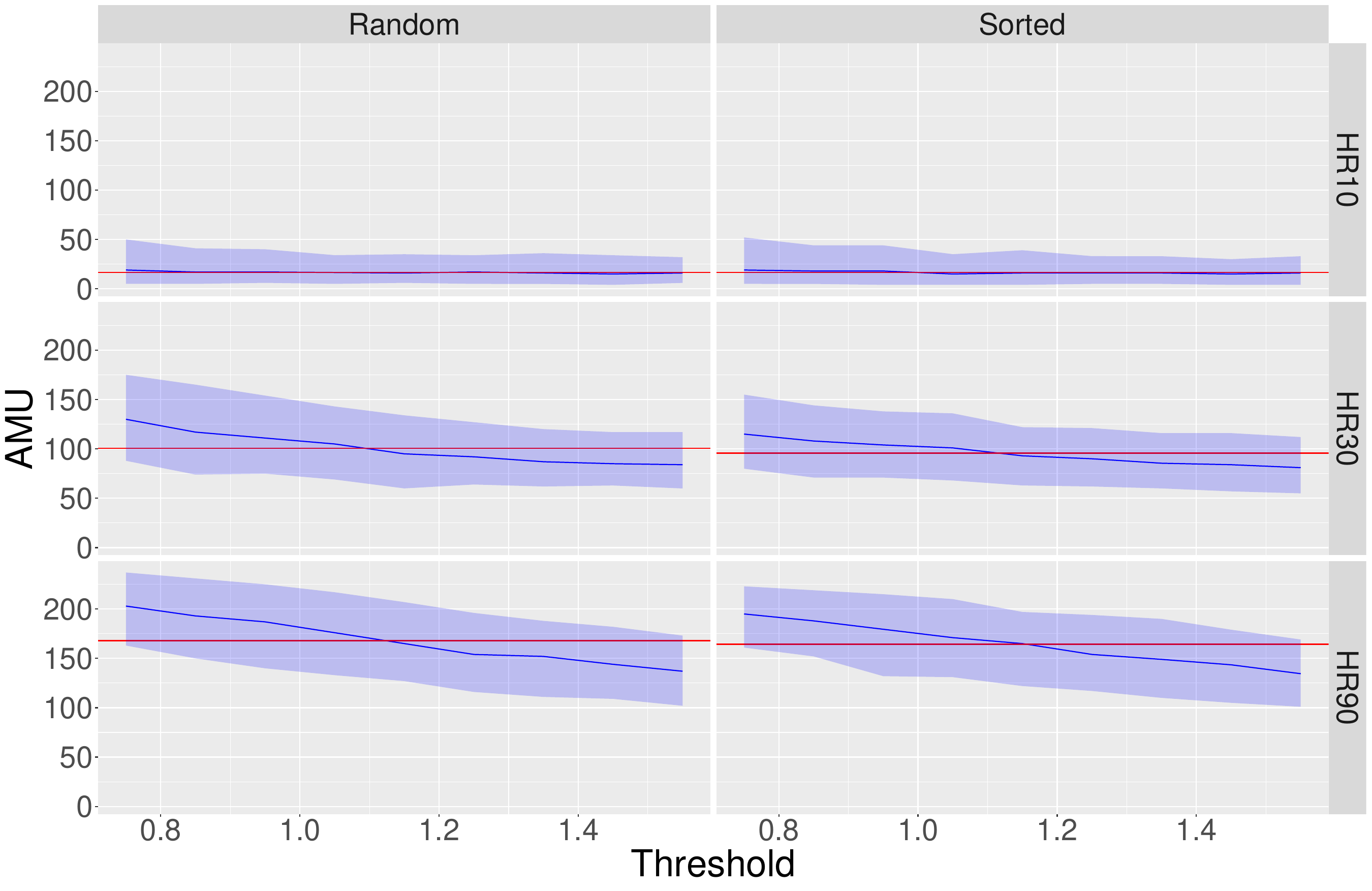}
    \caption{Median and 5-95\% quantiles of the AMU across scenarios and threshold values for the slope based criterion. The horizontal red line represents the mean value of the AMU accross the thresholds in the given scenario.}
    \label{5F}
\end{figure}

The AMU was also roughly constant in HR10 scenarios across the threshold values (Figure \ref{5F}). It decreased almost linearly in HR30 and HR90 scenarios. The AMU was lower than the cutoff in every scenario when the threshold for the slope criterion was higher than 1.15. We therefore chose this value as threshold for the slope criterion.
\newpage

\textbf{Threshold for the slope criterion: 1.15}\\
\textbf{Threshold for the severity criterion: 1.5}

\newpage
\section{Additional figures}

\begin{figure}[h!]
    \centering
    \includegraphics[width=0.85\textwidth]{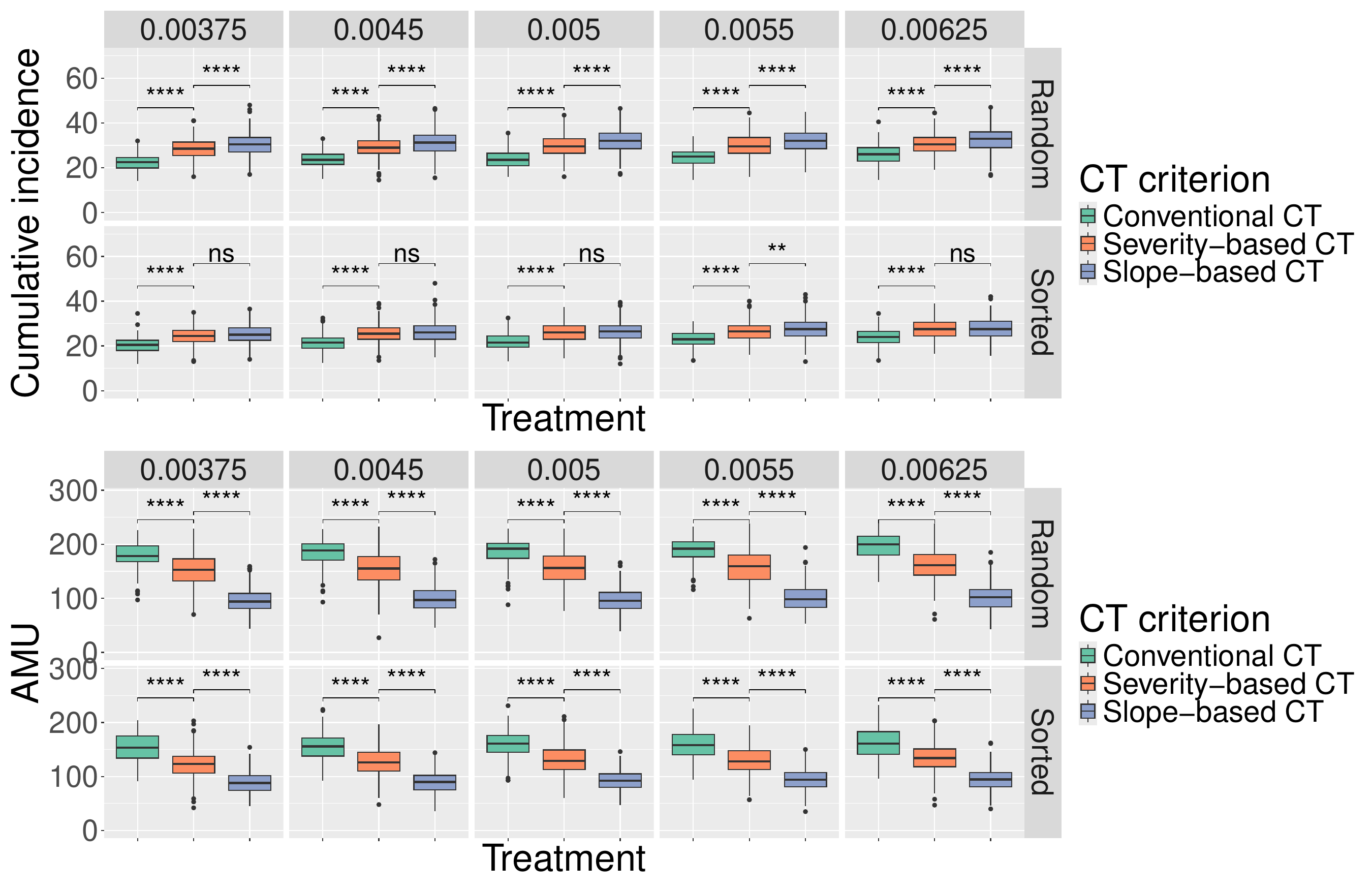}
    \caption{\textbf{Impact of the collective treatment strategies on the outputs with the tested  transmission rate values.} Top panel: Cumulative incidence. Bottom panel: AMU. First row: the animals are allocated randomly in pens. Second row: the animals are allocated in pens according to their individual risk level. Each column represents the tested value for the transmission rate. Statistical test: Wilcoxonn-Mann-Whitney test with Bonferroni's correction.}
    \label{sa_rate}
\end{figure}

\begin{figure}[h!]
    \centering
    \includegraphics[width=0.85\textwidth]{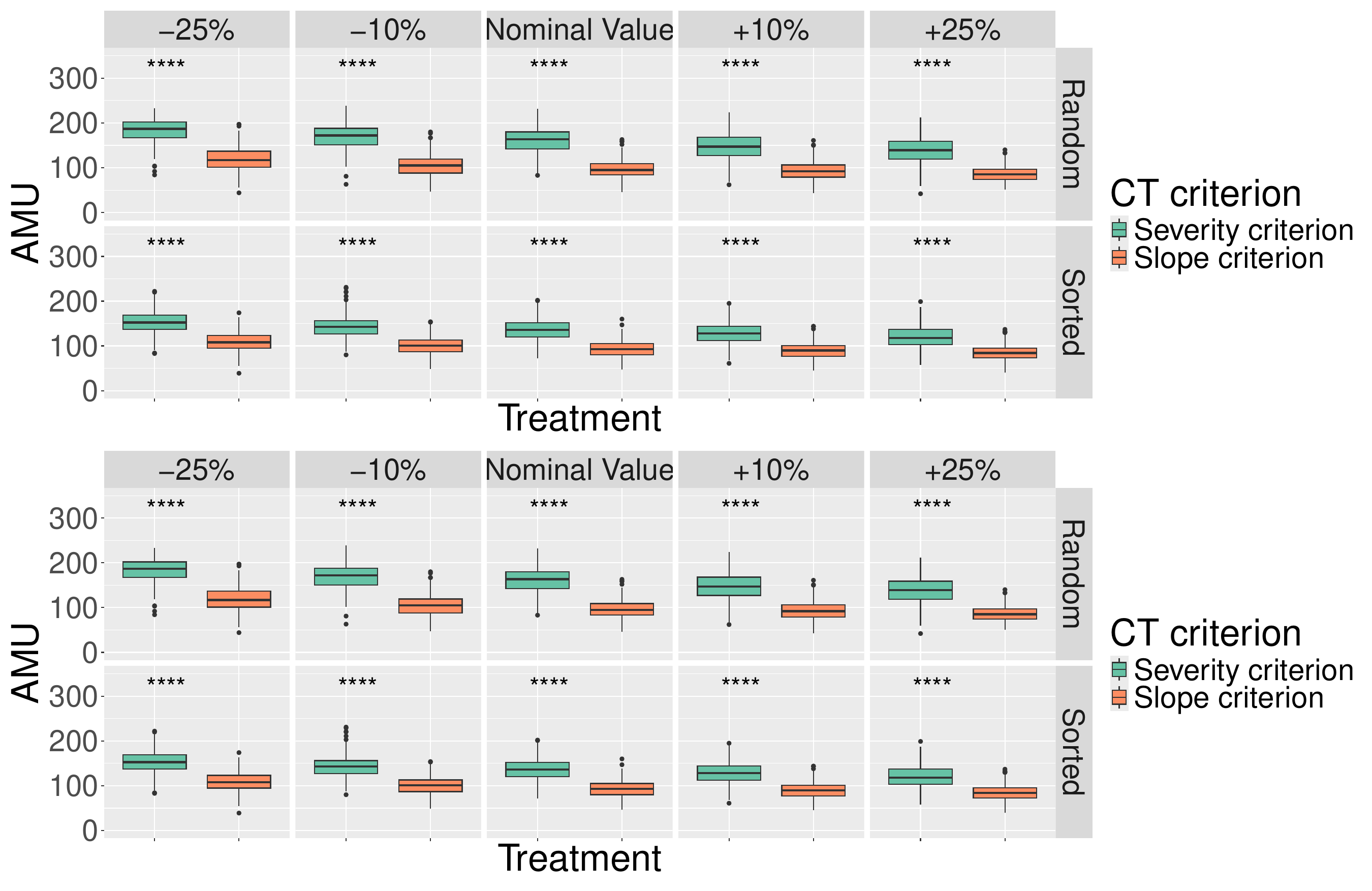}
    \caption{\textbf{Impact of the collective treatment strategies on the outputs with the tested  thresholds.} Top panel: Cumulative incidence. Bottom panel: AMU. First row: the animals are allocated randomly in pens. Second row: the animals are allocated in pens according to their individual risk level. Each column represents the tested value for the transmission rate. Statistical test: Wilcoxonn-Mann-Whitney test with Bonferroni's correction.}
    \label{sa_threshold}
\end{figure}

\newpage

\printbibliography

@article{delabouglise_linking_2017,
	title = {Linking disease epidemiology and livestock productivity: {The} case of bovine respiratory disease in {France}},
	volume = {12},
	issn = {1932-6203},
	shorttitle = {Linking disease epidemiology and livestock productivity},
	url = {https://dx.plos.org/10.1371/journal.pone.0189090},
	doi = {10.1371/journal.pone.0189090},
	language = {en},
	number = {12},
	urldate = {2021-12-09},
	journal = {PLOS ONE},
	author = {Delabouglise, Alexis and James, Andrew and Valarcher, Jean-François and Hagglünd, Sara and Raboisson, Didier and Rushton, Jonathan},
	editor = {Browning, Glenn F.},
	month = dec,
	year = {2017},
	note = {Number: 12},
	pages = {e0189090},
}

@article{assie_exposure_2009,
	title = {Exposure to pathogens and incidence of respiratory disease in young bulls on their arrival at fattening operations in {France}},
	volume = {165},
	issn = {00424900},
	url = {http://doi.wiley.com/10.1136/vr.165.7.195},
	doi = {10.1136/vr.165.7.195},
	language = {en},
	number = {7},
	urldate = {2021-12-13},
	journal = {Veterinary Record},
	author = {Assié, S. and Seegers, H. and Makoschey, B. and Désiré-Bousquié, L. and Bareille, N.},
	month = aug,
	year = {2009},
	note = {Number: 7},
	pages = {195--199},
}

@article{grissett_structured_2015,
	title = {Structured {Literature} {Review} of {Responses} of {Cattle} to {Viral} and {Bacterial} {Pathogens} {Causing} {Bovine} {Respiratory} {Disease} {Complex}},
	volume = {29},
	issn = {08916640},
	url = {https://onlinelibrary.wiley.com/doi/10.1111/jvim.12597},
	doi = {10.1111/jvim.12597},
	language = {en},
	number = {3},
	urldate = {2021-12-16},
	journal = {Journal of Veterinary Internal Medicine},
	author = {Grissett, G.P. and White, B.J. and Larson, R.L.},
	month = may,
	year = {2015},
	note = {Number: 3},
	pages = {770--780},
}

@article{griffin_bovine_2010,
	title = {Bovine {Pasteurellosis} and {Other} {Bacterial} {Infections} of the {Respiratory} {Tract}},
	volume = {26},
	issn = {07490720},
	url = {https://linkinghub.elsevier.com/retrieve/pii/S074907200900108X},
	doi = {10.1016/j.cvfa.2009.10.010},
	language = {en},
	number = {1},
	urldate = {2021-12-16},
	journal = {Veterinary Clinics of North America: Food Animal Practice},
	author = {Griffin, Dee},
	month = mar,
	year = {2010},
	note = {Number: 1},
	pages = {57--71},
}

@article{noyes_mannheimia_2015,
	title = {Mannheimia haemolytica in feedlot cattle: prevalence of recovery and associations with antimicrobial use, resistance, and health outcomes},
	volume = {29},
	issn = {1939-1676},
	shorttitle = {Mannheimia haemolytica in feedlot cattle},
	doi = {10.1111/jvim.12547},
	language = {eng},
	number = {2},
	journal = {Journal of Veterinary Internal Medicine},
	author = {Noyes, N. R. and Benedict, K. M. and Gow, S. P. and Booker, C. W. and Hannon, S. J. and McAllister, T. A. and Morley, P. S.},
	month = apr,
	year = {2015},
	pmid = {25818224},
	pmcid = {PMC4895489},
	note = {Number: 2},
	pages = {705--713},
}

@article{timsit_transmission_2013,
	title = {Transmission dynamics of {Mannheimia} haemolytica in newly-received beef bulls at fattening operations},
	volume = {161},
	issn = {03781135},
	url = {https://linkinghub.elsevier.com/retrieve/pii/S0378113512004464},
	doi = {10.1016/j.vetmic.2012.07.044},
	language = {en},
	number = {3-4},
	urldate = {2022-01-13},
	journal = {Veterinary Microbiology},
	author = {Timsit, E. and Christensen, H. and Bareille, N. and Seegers, H. and Bisgaard, M. and Assié, S.},
	month = jan,
	year = {2013},
	note = {Number: 3-4},
	pages = {295--304},
}

@article{timsit_prevalence_2017,
	title = {Prevalence and antimicrobial susceptibility of {Mannheimia} haemolytica, {Pasteurella} multocida, and {Histophilus} somni isolated from the lower respiratory tract of healthy feedlot cattle and those diagnosed with bovine respiratory disease},
	volume = {208},
	issn = {03781135},
	url = {https://linkinghub.elsevier.com/retrieve/pii/S0378113517306521},
	doi = {10.1016/j.vetmic.2017.07.013},
	language = {en},
	urldate = {2022-01-13},
	journal = {Veterinary Microbiology},
	author = {Timsit, Edouard and Hallewell, Jennyka and Booker, Calvin and Tison, Nicolas and Amat, Samat and Alexander, Trevor W.},
	month = sep,
	year = {2017},
	pages = {118--125},
}

@article{sudaryatma_bovine_2018,
	title = {Bovine respiratory syncytial virus infection enhances {Pasteurella} multocida adherence on respiratory epithelial cells},
	volume = {220},
	issn = {1873-2542},
	url = {https://europepmc.org/articles/PMC7117154},
	doi = {10.1016/j.vetmic.2018.04.031},
	language = {eng},
	urldate = {2022-01-24},
	journal = {Veterinary microbiology},
	author = {Sudaryatma, Putu Eka and Nakamura, Kimika and Mekata, Hirohisa and Sekiguchi, Satoshi and Kubo, Meiko and Kobayashi, Ikuo and Subangkit, Mawar and Goto, Yoshitaka and Okabayashi, Tamaki},
	month = jul,
	year = {2018},
	pmid = {29885798},
	pmcid = {PMC7117154},
	pages = {33--38},
}

@article{caswell_mycoplasma_2010,
	title = {Mycoplasma bovis in {Respiratory} {Disease} of {Feedlot} {Cattle}},
	volume = {26},
	issn = {0749-0720, 1558-4240},
	url = {https://www.vetfood.theclinics.com/article/S0749-0720(10)00004-6/fulltext},
	doi = {10.1016/j.cvfa.2010.03.003},
	language = {English},
	number = {2},
	urldate = {2022-01-25},
	journal = {Veterinary Clinics: Food Animal Practice},
	author = {Caswell, Jeff L. and Bateman, Ken G. and Cai, Hugh Y. and Castillo-Alcala, Fernanda},
	month = jul,
	year = {2010},
	pmid = {20619190},
	note = {Number: 2
Publisher: Elsevier},
	pages = {365--379},
}

@article{frank_colonization_1986,
	title = {Colonization of the nasal passages of calves with {Pasteurella} haemolytica serotype 1 and regeneration of colonization after experimentally induced viral infection of the respiratory tract},
	volume = {47},
	issn = {0002-9645},
	language = {eng},
	number = {8},
	journal = {American Journal of Veterinary Research},
	author = {Frank, G. H. and Briggs, R. E. and Gillette, K. G.},
	month = aug,
	year = {1986},
	pmid = {3752678},
	note = {Number: 8},
	pages = {1704--1707},
}

@article{thomas_insights_2019,
	title = {Insights into {Pasteurellaceae} carriage dynamics in the nasal passages of healthy beef calves},
	volume = {9},
	issn = {2045-2322},
	url = {https://www.ncbi.nlm.nih.gov/pmc/articles/PMC6697682/},
	doi = {10.1038/s41598-019-48007-5},
	urldate = {2022-07-05},
	journal = {Scientific Reports},
	author = {Thomas, A. C. and Bailey, M. and Lee, M. R. F. and Mead, A. and Morales-Aza, B. and Reynolds, R. and Vipond, B. and Finn, A. and Eisler, M. C.},
	month = aug,
	year = {2019},
	pmid = {31420565},
	pmcid = {PMC6697682},
	pages = {11943},
}

@article{klima_characterization_2012,
	title = {Characterization of {Mannheimia} haemolytica isolated from feedlot cattle that were healthy or treated for bovine respiratory disease},
	language = {en},
	author = {Klima, Cassidy L and Alexander, Trevor W and Hendrick, Steve and McAllister, Tim A},
	year = {2012},
	pages = {8},
}

@article{dedonder_review_2015,
	series = {Bovine {Clinical} {Pharmacology}},
	title = {A {Review} of the {Expected} {Effects} of {Antimicrobials} in {Bovine} {Respiratory} {Disease} {Treatment} and {Control} {Using} {Outcomes} from {Published} {Randomized} {Clinical} {Trials} with {Negative} {Controls}},
	volume = {31},
	issn = {0749-0720},
	url = {https://www.sciencedirect.com/science/article/pii/S0749072014000838},
	doi = {10.1016/j.cvfa.2014.11.003},
	language = {en},
	number = {1},
	urldate = {2022-07-19},
	journal = {Veterinary Clinics of North America: Food Animal Practice},
	author = {DeDonder, Keith D. and Apley, Michael D.},
	month = mar,
	year = {2015},
	note = {Number: 1},
	pages = {97--111},
}

@article{hilton_brd_2014,
	title = {{BRD} in 2014: where have we been, where are we now, and where do we want to go?},
	volume = {15},
	issn = {1466-2523, 1475-2654},
	shorttitle = {{BRD} in 2014},
	url = {https://www.cambridge.org/core/journals/animal-health-research-reviews/article/abs/brd-in-2014-where-have-we-been-where-are-we-now-and-where-do-we-want-to-go/B9B8135CE1212B412B5166B7DEC11F5C},
	doi = {10.1017/S1466252314000115},
	language = {en},
	number = {2},
	urldate = {2022-07-21},
	journal = {Animal Health Research Reviews},
	author = {Hilton, W. Mark},
	month = dec,
	year = {2014},
	note = {Number: 2
Publisher: Cambridge University Press},
	pages = {120--122},
}

@article{gaudino_understanding_2022,
	title = {Understanding the mechanisms of viral and bacterial coinfections in bovine respiratory disease: a comprehensive literature review of experimental evidence},
	volume = {53},
	issn = {1297-9716},
	shorttitle = {Understanding the mechanisms of viral and bacterial coinfections in bovine respiratory disease},
	url = {https://doi.org/10.1186/s13567-022-01086-1},
	doi = {10.1186/s13567-022-01086-1},
	number = {1},
	urldate = {2022-09-19},
	journal = {Veterinary Research},
	author = {Gaudino, Maria and Nagamine, Brandy and Ducatez, Mariette F. and Meyer, Gilles},
	month = sep,
	year = {2022},
	note = {Number: 1},
	pages = {70},
}

@article{picault_emulsion_2019,
	title = {{EMULSION}: transparent and flexible multiscale stochastic models in human, animal and plant epidemiology},
	volume = {15},
	number = {9},
	journal = {PLoS computational biology},
	author = {Picault, Sébastien and Huang, Yu-Lin and Sicard, Vianney and Arnoux, Sandie and Beaunée, Gaël and Ezanno, Pauline},
	year = {2019},
	note = {Number: 9
Publisher: Public Library of Science San Francisco, CA USA},
	pages = {e1007342},
}

@article{ollivett_brd_2020,
	title = {{BRD} treatment failure: clinical and pathologic considerations},
	volume = {21},
	number = {2},
	journal = {Animal Health Research Reviews},
	author = {Ollivett, TL},
	year = {2020},
	note = {Number: 2
Publisher: Cambridge University Press},
	pages = {175--176},
}

@article{laxminarayan_antibiotic_2013,
	title = {Antibiotic resistance—the need for global solutions},
	volume = {13},
	number = {12},
	journal = {The Lancet infectious diseases},
	author = {Laxminarayan, Ramanan and Duse, Adriano and Wattal, Chand and Zaidi, Anita KM and Wertheim, Heiman FL and Sumpradit, Nithima and Vlieghe, Erika and Hara, Gabriel Levy and Gould, Ian M and Goossens, Herman and {others}},
	year = {2013},
	note = {Number: 12
Publisher: Elsevier},
	pages = {1057--1098},
}

@article{rice_mannheimia_2007,
	title = {Mannheimia haemolytica and bovine respiratory disease},
	volume = {8},
	number = {2},
	journal = {Animal Health Research Reviews},
	author = {Rice, JA and Carrasco-Medina, L and Hodgins, DC and Shewen, PE},
	year = {2007},
	note = {Number: 2
Publisher: Cambridge University Press},
	pages = {117--128},
}

@article{griffin_monster_2014,
	title = {The monster we don't see: subclinical {BRD} in beef cattle},
	volume = {15},
	number = {2},
	journal = {Animal health research reviews},
	author = {Griffin, Dee},
	year = {2014},
	note = {Number: 2
Publisher: Cambridge University Press},
	pages = {138--141},
}

@article{coetzee_association_2019,
	title = {Association between antimicrobial drug class for treatment and retreatment of bovine respiratory disease ({BRD}) and frequency of resistant {BRD} pathogen isolation from veterinary diagnostic laboratory samples},
	volume = {14},
	number = {12},
	journal = {PloS one},
	author = {Coetzee, Johann F and Magstadt, Drew R and Sidhu, Pritam K and Follett, Lendie and Schuler, Adlai M and Krull, Adam C and Cooper, Vickie L and Engelken, Terry J and Kleinhenz, Michael D and O’Connor, Annette M},
	year = {2019},
	note = {Number: 12
Publisher: Public Library of Science San Francisco, CA USA},
	pages = {e0219104},
}

@article{amrine_evaluation_2019,
	title = {Evaluation of three classification models to predict risk class of cattle cohorts developing bovine respiratory disease within the first 14 days on feed using on-arrival and/or pre-arrival information},
	volume = {156},
	issn = {0168-1699},
	url = {https://www.sciencedirect.com/science/article/pii/S0168169918304642},
	doi = {10.1016/j.compag.2018.11.035},
	language = {en},
	urldate = {2022-10-17},
	journal = {Computers and Electronics in Agriculture},
	author = {Amrine, David E. and McLellan, Jiena G. and White, Brad J. and Larson, Robert L. and Renter, David G. and Sanderson, Mike},
	month = jan,
	year = {2019},
	pages = {439--446},
}

@article{picault_modelling_2022,
	title = {Modelling the effects of antimicrobial metaphylaxis and pen size on bovine respiratory disease in high and low risk fattening cattle},
	volume = {53},
	issn = {1297-9716},
	url = {https://doi.org/10.1186/s13567-022-01094-1},
	doi = {10.1186/s13567-022-01094-1},
	number = {1},
	urldate = {2022-10-24},
	journal = {Veterinary Research},
	author = {Picault, Sébastien and Ezanno, Pauline and Smith, Kristen and Amrine, David and White, Brad and Assié, Sébastien},
	month = oct,
	year = {2022},
	note = {Number: 1},
	pages = {77},
}

@article{kudirkiene_occurrence_2021,
	title = {Occurrence of major and minor pathogens in calves diagnosed with bovine respiratory disease},
	volume = {259},
	issn = {0378-1135},
	url = {https://www.sciencedirect.com/science/article/pii/S0378113521001589},
	doi = {10.1016/j.vetmic.2021.109135},
	language = {en},
	urldate = {2022-12-14},
	journal = {Veterinary Microbiology},
	author = {Kudirkiene, Egle and Aagaard, Anne Katrine and Schmidt, Louise M. B. and Pansri, Potjamas and Krogh, Kenneth M. and Olsen, John E.},
	month = aug,
	year = {2021},
	pages = {109135},
}

@article{ives_use_2015,
	title = {Use of {Antimicrobial} {Metaphylaxis} for the {Control} of {Bovine} {Respiratory} {Disease} in {High}-{Risk} {Cattle}},
	volume = {31},
	issn = {1558-4240},
	doi = {10.1016/j.cvfa.2015.05.008},
	language = {eng},
	number = {3},
	journal = {The Veterinary Clinics of North America. Food Animal Practice},
	author = {Ives, Samuel E. and Richeson, John T.},
	month = nov,
	year = {2015},
	pmid = {26227871},
	pages = {341--350, v},
}

@misc{noauthor_migale_nodate,
	title = {Migale platform {\textbar} {Migale}},
	url = {https://migale.inrae.fr/},
	urldate = {2023-02-02},
}

@article{nickell_metaphylactic_2010,
	series = {Bovine {Respiratory} {Disease}},
	title = {Metaphylactic {Antimicrobial} {Therapy} for {Bovine} {Respiratory} {Disease} in {Stocker} and {Feedlot} {Cattle}},
	volume = {26},
	issn = {0749-0720},
	url = {https://www.sciencedirect.com/science/article/pii/S0749072010000125},
	doi = {10.1016/j.cvfa.2010.04.006},
	language = {en},
	number = {2},
	urldate = {2023-04-25},
	journal = {Veterinary Clinics of North America: Food Animal Practice},
	author = {Nickell, Jason S. and White, Brad J.},
	month = jul,
	year = {2010},
	pages = {285--301},
}

@article{kayser_evaluation_2019,
	title = {Evaluation of statistical process control procedures to monitor feeding behavior patterns and detect onset of bovine respiratory disease in growing bulls},
	volume = {97},
	issn = {0021-8812},
	url = {https://www.ncbi.nlm.nih.gov/pmc/articles/PMC6396243/},
	doi = {10.1093/jas/sky486},
	number = {3},
	urldate = {2023-05-22},
	journal = {Journal of Animal Science},
	author = {Kayser, William C and Carstens, Gordon E and Jackson, Kirby S and Pinchak, William E and Banerjee, Amarnath and Fu, Yu},
	month = mar,
	year = {2019},
	pmid = {30590611},
	pmcid = {PMC6396243},
	pages = {1158--1170},
}

@article{blakebrough-hall_factors_2022,
	title = {Factors associated with bovine respiratory disease case fatality in feedlot cattle},
	volume = {100},
	issn = {1525-3163},
	url = {https://doi.org/10.1093/jas/skab361},
	doi = {10.1093/jas/skab361},
	number = {1},
	urldate = {2023-05-24},
	journal = {Journal of Animal Science},
	author = {Blakebrough-Hall, Claudia and Hick, Paul and Mahony, T J and González, Luciano A},
	month = jan,
	year = {2022},
	pages = {skab361},
}

@article{edwards_control_2010,
	title = {Control {Methods} for {Bovine} {Respiratory} {Disease} for {Feedlot} {Cattle}},
	volume = {26},
	issn = {07490720},
	url = {https://linkinghub.elsevier.com/retrieve/pii/S074907201000006X},
	doi = {10.1016/j.cvfa.2010.03.005},
	language = {en},
	number = {2},
	urldate = {2023-07-03},
	journal = {Veterinary Clinics of North America: Food Animal Practice},
	author = {Edwards, T.A.},
	month = jul,
	year = {2010},
	pages = {273--284},
}

@article{sorin-dupont_modeling_2023,
	title = {Modeling the effects of farming practices on bovine respiratory disease in a multi-batch cattle fattening farm},
	volume = {219},
	issn = {0167-5877},
	url = {https://www.sciencedirect.com/science/article/pii/S0167587723001733},
	doi = {10.1016/j.prevetmed.2023.106009},
	urldate = {2023-10-18},
	journal = {Preventive Veterinary Medicine},
	author = {Sorin-Dupont, Baptiste and Picault, Sebastien and Pardon, Bart and Ezanno, Pauline and Assié, Sebastien},
	month = oct,
	year = {2023},
	pages = {106009},
}

@article{woolums_multidrug_2018,
	title = {Multidrug resistant {Mannheimia} haemolytica isolated from high-risk beef stocker cattle after antimicrobial metaphylaxis and treatment for bovine respiratory disease},
	volume = {221},
	issn = {0378-1135},
	url = {https://www.sciencedirect.com/science/article/pii/S037811351730977X},
	doi = {10.1016/j.vetmic.2018.06.005},
	urldate = {2023-12-01},
	journal = {Veterinary Microbiology},
	author = {Woolums, Amelia R. and Karisch, Brandi B. and Frye, Jonathan G. and Epperson, William and Smith, David R. and Blanton, John and Austin, Frank and Kaplan, Ray and Hiott, Lari and Woodley, Tiffanie and Gupta, Sushim K. and Jackson, Charlene R. and McClelland, Michael},
	month = jul,
	year = {2018},
	pages = {143--152},
}

@article{abell_mixed_2017,
	title = {A mixed treatment comparison meta-analysis of metaphylaxis treatments for bovine respiratory disease in beef cattle},
	volume = {95},
	issn = {1525-3163},
	doi = {10.2527/jas.2016.1062},
	language = {eng},
	number = {2},
	journal = {Journal of Animal Science},
	author = {Abell, K. M. and Theurer, M. E. and Larson, R. L. and White, B. J. and Apley, M.},
	month = feb,
	year = {2017},
	pmid = {28380607},
	pages = {626--635},
}

@article{oconnor_mixed_2013,
	title = {A mixed treatment comparison meta-analysis of antibiotic treatments for bovine respiratory disease},
	volume = {110},
	issn = {0167-5877},
	url = {https://www.sciencedirect.com/science/article/pii/S0167587712004023},
	doi = {10.1016/j.prevetmed.2012.11.025},
	number = {2},
	urldate = {2023-12-01},
	journal = {Preventive Veterinary Medicine},
	author = {O’Connor, Annette M. and Coetzee, Johann F. and da Silva, Natalia and Wang, Chong},
	month = jun,
	year = {2013},
	pages = {77--87},
}

@article{chen_scoping_2022,
	title = {Scoping {Review} on {Risk} {Factors} and {Methods} for the {Prevention} of {Bovine} {Respiratory} {Disease} {Applicable} to {Cow}–{Calf} {Operations}},
	volume = {12},
	copyright = {http://creativecommons.org/licenses/by/3.0/},
	issn = {2076-2615},
	url = {https://www.mdpi.com/2076-2615/12/3/334},
	doi = {10.3390/ani12030334},
	language = {en},
	number = {3},
	urldate = {2023-12-01},
	journal = {Animals},
	author = {Chen, Shih-Yu and Negri Bernardino, Pedro and Fausak, Erik and Van Noord, Megan and Maier, Gabriele},
	month = jan,
	year = {2022},
	note = {Number: 3
Publisher: Multidisciplinary Digital Publishing Institute},
	pages = {334},
}

@article{gonzalez-martin_reducing_2011,
	title = {Reducing antibiotic use: {Selective} metaphylaxis with florfenicol in commercial feedlots},
	volume = {141},
	issn = {1871-1413},
	shorttitle = {Reducing antibiotic use},
	url = {https://www.sciencedirect.com/science/article/pii/S1871141311002046},
	doi = {10.1016/j.livsci.2011.05.016},
	number = {2},
	urldate = {2024-02-02},
	journal = {Livestock Science},
	author = {González-Martín, J. V. and Elvira, L. and Cerviño López, M. and Pérez Villalobos, N. and Calvo López-Guerrero, E. and Astiz, S.},
	month = nov,
	year = {2011},
	pages = {173--181},
}

@article{baptiste_antimicrobial_2017,
	title = {Do antimicrobial mass medications work? {A} systematic review and meta-analysis of randomised clinical trials investigating antimicrobial prophylaxis or metaphylaxis against naturally occurring bovine respiratory disease},
	volume = {75},
	issn = {2049-632X},
	shorttitle = {Do antimicrobial mass medications work?},
	url = {https://doi.org/10.1093/femspd/ftx083},
	doi = {10.1093/femspd/ftx083},
	number = {7},
	urldate = {2024-02-02},
	journal = {Pathogens and Disease},
	author = {Baptiste, Keith Edward and Kyvsgaard, Niels Christian},
	month = sep,
	year = {2017},
	pages = {ftx083},
}

@article{lees_rational_2002,
	title = {Rational dosing of antimicrobial drugs: animals versus humans},
	volume = {19},
	issn = {0924-8579},
	shorttitle = {Rational dosing of antimicrobial drugs},
	doi = {10.1016/s0924-8579(02)00025-0},
	language = {eng},
	number = {4},
	journal = {International Journal of Antimicrobial Agents},
	author = {Lees, Peter and Shojaee Aliabadi, Fariborz},
	month = apr,
	year = {2002},
	pmid = {11978498},
	pages = {269--284},
}

@book{radostits_herd_2001,
	address = {Philadelphia},
	edition = {3rd ed},
	title = {Herd health : food animal production medicine},
	isbn = {978-0-7216-7694-4 0-7216-7694-4},
	language = {eng},
	publisher = {Saunders Philadelphia},
	author = {Radostits, O. M.},
	year = {2001},
	note = {Section: xi, 884 pages : illustrations ; 29 cm},
}

@book{mornet_p_espinasse_j_veau_1977,
	title = {Le {Veau} : {Anatomie}, physiologie, élevage, alimentation, production},
	author = {Mornet P., Espinasse J.},
	year = {1977},
}

@article{ezanno_how_2020,
	title = {How mechanistic modelling supports decision making for the control of enzootic infectious diseases},
	volume = {32},
	issn = {1755-4365},
	url = {https://www.sciencedirect.com/science/article/pii/S1755436520300256},
	doi = {10.1016/j.epidem.2020.100398},
	urldate = {2024-02-07},
	journal = {Epidemics},
	author = {Ezanno, P. and Andraud, M. and Beaunée, G. and Hoch, T. and Krebs, S. and Rault, A. and Touzeau, S. and Vergu, E. and Widgren, S.},
	month = sep,
	year = {2020},
	pages = {100398},
}

@article{booker_microbiological_2008,
	title = {Microbiological and histopathological findings in cases of fatal bovine respiratory disease of feedlot cattle in western {Canada}},
	volume = {49},
	issn = {0008-5286},
	url = {https://www.ncbi.nlm.nih.gov/pmc/articles/PMC2359492/},
	number = {5},
	urldate = {2024-03-28},
	journal = {The Canadian Veterinary Journal},
	author = {Booker, Calvin W. and Abutarbush, Sameeh M. and Morley, Paul S. and Jim, G. Kee and Pittman, Tom J. and Schunicht, Oliver C. and Perrett, Tye and Wildman, Brian K. and Fenton, R. Kent and Guichon, P. Timothy and Janzen, Eugene D.},
	month = may,
	year = {2008},
	pmid = {18512458},
	pmcid = {PMC2359492},
	pages = {473--481},
}

@article{welsh_isolation_2004,
	title = {Isolation and antimicrobial susceptibilities of bacterial pathogens from bovine pneumonia: 1994--2002},
	volume = {16},
	issn = {1040-6387},
	shorttitle = {Isolation and antimicrobial susceptibilities of bacterial pathogens from bovine pneumonia},
	doi = {10.1177/104063870401600510},
	language = {eng},
	number = {5},
	journal = {Journal of Veterinary Diagnostic Investigation: Official Publication of the American Association of Veterinary Laboratory Diagnosticians, Inc},
	author = {Welsh, Ronald D. and Dye, Laura B. and Payton, Mark E. and Confer, Anthony W.},
	month = sep,
	year = {2004},
	pmid = {15460326},
	pages = {426--431},
}

@article{terry_strategies_2021,
	title = {Strategies to improve the efficiency of beef cattle production},
	volume = {101},
	issn = {0008-3984},
	url = {https://cdnsciencepub.com/doi/full/10.1139/cjas-2020-0022},
	doi = {10.1139/cjas-2020-0022},
	number = {1},
	urldate = {2024-04-29},
	journal = {Canadian Journal of Animal Science},
	author = {Terry, Stephanie A. and Basarab, John A. and Guan, Le Luo and McAllister, Tim A.},
	month = mar,
	year = {2021},
	note = {Publisher: NRC Research Press},
	pages = {1--19},
}

@article{watts_antimicrobial_2010,
	title = {Antimicrobial {Resistance} in {Bovine} {Respiratory} {Disease} {Pathogens}: {Measures}, {Trends}, and {Impact} on {Efficacy}},
	volume = {26},
	copyright = {https://www.elsevier.com/tdm/userlicense/1.0/},
	issn = {07490720},
	shorttitle = {Antimicrobial {Resistance} in {Bovine} {Respiratory} {Disease} {Pathogens}},
	url = {https://linkinghub.elsevier.com/retrieve/pii/S0749072009001078},
	doi = {10.1016/j.cvfa.2009.10.009},
	language = {en},
	number = {1},
	urldate = {2024-05-13},
	journal = {Veterinary Clinics of North America: Food Animal Practice},
	author = {Watts, Jeffrey L. and Sweeney, Michael T.},
	month = mar,
	year = {2010},
	pages = {79--88},
}

@article{bateman_evaluation_1990,
	title = {An evaluation of antimicrobial therapy for undifferentiated bovine respiratory disease},
	volume = {31},
	issn = {0008-5286},
	url = {https://www.ncbi.nlm.nih.gov/pmc/articles/PMC1480842/},
	number = {10},
	urldate = {2024-05-13},
	journal = {The Canadian Veterinary Journal},
	author = {Bateman, Ken G. and Martin, S. Wayne and Shewen, Patricia E. and Menzies, Paula I.},
	month = oct,
	year = {1990},
	pmid = {17423676},
	pmcid = {PMC1480842},
	pages = {689--696},
}

@article{vanbergue_comparison_2024,
	title = {Comparison between a complete preconditioning programme and conventional conduct on behaviour, health and performance of young bulls from small cow-calf herds},
	volume = {18},
	issn = {1751-7311},
	url = {https://www.sciencedirect.com/science/article/pii/S1751731124001009},
	doi = {10.1016/j.animal.2024.101169},
	number = {6},
	urldate = {2024-05-28},
	journal = {animal},
	author = {Vanbergue, E. and Assie, S. and Mounaix, B. and Guiadeur, M. and Robert, F. and Andrieu, D. and Cebron, N. and Meyer, G. and Philibert, A. and Foucras, G.},
	month = jun,
	year = {2024},
	pages = {101169},
}

@article{fulton_lung_2009,
	title = {Lung {Pathology} and {Infectious} {Agents} in {Fatal} {Feedlot} {Pneumonias} and {Relationship} with {Mortality}, {Disease} {Onset}, and {Treatments}},
	volume = {21},
	issn = {1040-6387},
	url = {https://doi.org/10.1177/104063870902100407},
	doi = {10.1177/104063870902100407},
	language = {en},
	number = {4},
	urldate = {2024-06-13},
	journal = {Journal of Veterinary Diagnostic Investigation},
	author = {Fulton, Robert W. and Blood, K. Shawn and Panciera, Roger J. and Payton, Mark E. and Ridpath, Julia F. and Confer, Anthony W. and Saliki, Jeremiah T. and Burge, Lurinda T. and Welsh, Ronald D. and Johnson, Bill J. and Reck, Amy},
	month = jul,
	year = {2009},
	note = {Publisher: SAGE Publications Inc},
	pages = {464--477},
}

@article{baggott2011demonstration,
  title={Demonstration of the metaphylactic use of gamithromycin against bacterial pathogens associated with bovine respiratory disease in a multicentre farm trial},
  author={Baggott, D and Casartelli, A and Fraisse, F and Manavella, C and Marteau, R and Rehbein, S and Wiedemann, M and Yoon, S},
  journal={Veterinary Record},
  volume={168},
  number={9},
  pages={241--241},
  year={2011},
  publisher={Wiley Online Library}
}

@article{kamel2024strategies,
  title={Strategies for Bovine Respiratory Disease (BRD) Diagnosis and Prognosis: A Comprehensive Overview},
  author={Kamel, Mohamed S and Davidson, Josiah Levi and Verma, Mohit S},
  journal={Animals},
  volume={14},
  number={4},
  pages={627},
  year={2024},
  publisher={MDPI}
}

@article{van2017reducing,
  title={Reducing antimicrobial use in food animals},
  author={Van Boeckel, Thomas P and Glennon, Emma E and Chen, Dora and Gilbert, Marius and Robinson, Timothy P and Grenfell, Bryan T and Levin, Simon A and Bonhoeffer, Sebastian and Laxminarayan, Ramanan},
  journal={Science},
  volume={357},
  number={6358},
  pages={1350--1352},
  year={2017},
  publisher={American Association for the Advancement of Science}
}

@article{ackermann2000response,
  title={Response of the ruminant respiratory tract to Mannheimia (Pasteurella) haemolytica},
  author={Ackermann, Mark R and Brogden, Kim A},
  journal={Microbes and infection},
  volume={2},
  number={9},
  pages={1079--1088},
  year={2000},
  publisher={Elsevier}
}

@article{timsit2016diagnostic,
  title={Diagnostic accuracy of clinical illness for bovine respiratory disease (BRD) diagnosis in beef cattle placed in feedlots: a systematic literature review and hierarchical Bayesian latent-class meta-analysis},
  author={Timsit, E and Dendukuri, N and Schiller, I and Buczinski, S},
  journal={Preventive Veterinary Medicine},
  volume={135},
  pages={67--73},
  year={2016},
  publisher={Elsevier}
}
\end{document}